\documentclass[a4paper,11pt]{article}
\pdfoutput=1

\usepackage{jheppub}
\usepackage[T1]{fontenc}
\usepackage{tensor}
\usepackage{subcaption}
\hypersetup{bookmarksnumbered}
\usepackage{bbold}
\usepackage{float}
\usepackage{mathtools,slashed}
\usepackage[table,dvipsnames]{xcolor}
\usepackage{longtable}
\usepackage[labelformat=simple]{subcaption}

\usepackage[colorinlistoftodos]{todonotes}
\usepackage{dsfont}

\usepackage[smalltableaux]{ytableau}

\DeclareFontFamily{OT1}{pzc}{}
\DeclareFontShape{OT1}{pzc}{m}{it}{<-> s * [1.10] pzcmi7t}{}
\DeclareMathAlphabet{\mathpzc}{OT1}{pzc}{m}{it}

\def\avg#1{\left<#1\right>}
\def\tr#1{\text{Tr}\left(#1\right)}

\title{\boldmath Bootstrapping Pion Form Factors at Large $N$}

\author{Jan Albert,$^\gamma$}
\author{Dilara Kosva$^\pi$}
\author{and Leonardo Rastelli$^\pi$}

\affiliation{$^\gamma$Princeton Center for Theoretical Science, Princeton University,\newline Princeton, NJ 08544-0708, U.S.A.}

\affiliation{$^\pi$C. N. Yang Institute for Theoretical Physics, Stony Brook University,\newline Stony Brook, NY 11794-3840, U.S.A.}

\abstract{
We initiate a bootstrap study of pion form factors in large $N$ QCD. We consider the mixed system of the vector-current two-point function, the pion vector form factor, and the pion scattering amplitude in the chiral limit. At large $N$ these observables are meromorphic, with  spectral data constrained by unitarity, crossing symmetry, and Regge boundedness. 
We obtain bounds of two
kinds. The first are rigorous and universal: from analyticity, unitarity
and the asymptotic Brodsky-Farrar scaling, we constrain low-energy form-factor coefficients. The second are more phenomenological, of the
Shifman-Vainshtein-Zakharov type: feeding in the perturbative ultraviolet
behavior at a finite scale lets us bound the pion decay constant, convert a large
$N$ lattice measurement into a lower bound on the scale at which asymptotic
freedom sets in, and constrain the pion charge radius. Combining these inputs,
the space of allowed chiral Lagrangians shrinks toward the region where large~$N$ QCD is
expected to sit. Our results illustrate how local gauge-invariant probes provide a canonical bridge between the hadronic bootstrap and the microscopic QCD Lagrangian.
}

\begin{document} 
\maketitle

\section{Introduction and summary}

The program of carving out the space of large $N$ confining gauge theories by modern bootstrap methods was initiated in \cite{Albert:2022oes} and developed in \cite{Albert:2023jtd,Albert:2023seb,Fernandez:2022kzi,Ma:2023vgc,Li:2023qzs,Eckner:2024pqt,Berman:2024kdh}, with the ultimate goal of cornering large $N$ QCD. The rules of the game are by now familiar. At $N=\infty$, mesons are infinitely narrow, their $2\to2$ scattering amplitudes are meromorphic functions of the Mandelstam invariants, and the bootstrap problem -- carving out the space of spectra and on-shell couplings compatible with crossing, unitarity and Regge boundedness -- is rigorously posed. Semidefinite programming methods,\footnote{Positivity bounds date back to \cite{Martin1969,Pham:1985cr,Ananthanarayan:1994hf,Pennington:1994kc,Comellas:1995hq,Dita:1998mh,Adams:2006sv}, but they were only systematized with the identification of null constraints in \cite{Tolley:2020gtv, Caron-Huot:2020cmc, Arkani-Hamed:2020blm}. Since then, they have experienced an explosion of activity with applications in a plethora of examples, aside from large $N$ gauge theories. See \cite{Caron_Huot_2021,Caron-Huot:2021enk,Bern_2021,Caron-Huot:2022ugt,Caron-Huot:2022jli,Henriksson_2022light,Henriksson_2022weakgravity,Berman:2023jys, McPeak:2023wmq,Eckner:2024ggx,Bertucci:2024qzt,Berman:2024wyt,Berman:2024eid,Albert:2024yap,H_ring_2024,Dong:2024omo,Berman:2025owb,Bellazzini:2025shd,Bellazzini:2025bay,Elvang:2026pmc,Boisvert:2026sfh} for a necessarily incomplete list.} imported from the conformal bootstrap, then yield two-sided bounds on the Wilson coefficients of the chiral Lagrangian, in units of the mass of the lightest meson exchanged in pion scattering, the rho.

A standing challenge for this program is that the resulting bounds are universal. They follow from general principles, supplemented at most by simple spectral assumptions, and apply verbatim to any large $N$ confining theory with the standard pattern of chiral symmetry breaking. They know about large $N$ kinematics, but they are {\it a priori} completely blind to the defining microscopic dynamics of the underlying gauge theory.

There are two complementary strategies that one can pursue to corner large $N$ QCD. The most conceptually straightforward (but technically very challenging) is to enlarge the set of $2 \to 2$ scattering processes, to include the lightest massive mesons as external states. The first step in this program, currently being finalized in~\cite{rhos}, is to consider the $2 \to 2$ scattering amplitude of the rho meson. The resulting bootstrap program is significantly more constrained, and one might hope that large $N$ QCD  emerges as a distinguished extremal solution at the exclusion boundary. This is not far fetched,
as already in the pion bootstrap the minimal spectral assumption that the first exchanged resonances are the rho and the $f_2$ mesons (with prescribed mass ratio) led to an extremal solution~\cite{Albert:2023seb} comprising a (single) Regge trajectory, whose low-lying states  are in uncanny agreement with the real world.

The second strategy is to  feed quantitative information of the microscopic Lagrangian into the dispersive machinery.
Why is this difficult? The high-energy limit relevant for dispersion relations of $2\to 2$ pion amplitudes is the Regge limit, $|s|\to \infty$ at fixed momentum transfer. This limit is controlled by the leading Regge trajectory -- intrinsically hadronic data, not directly computable from the QCD Lagrangian. The high-energy regime that \emph{is} dictated by asymptotic freedom is instead the fixed-angle limit, $|s|\to\infty$ with $u/s$ fixed, famously captured by the Brodsky-Farrar counting rules \cite{Brodsky:1973kr,Lepage:1980fj}. Developing a bootstrap setup probing the fixed-angle behavior is an interesting strategy in its own right, see \cite{Bocchia:2026kew} for a recent exploration of this regime with primal methods.

In this paper we follow what we regard as the canonical route for importing the short-distance physics of QCD into the hadronic bootstrap: correlation functions of \emph{local gauge-invariant operators}. The statement that QCD is described at short distances by weakly coupled quarks and gluons is most directly expressed through the operator product expansion, which puts the high-energy behavior of such correlators under perturbative control \cite{Wilson:1969zs,Gross:1973id,Politzer:1973fx}.\footnote{Another instance where this becomes apparent is in the fusion of parallel Wilson lines, which controls the UV behavior of confining strings \cite{Albert:2026fqj}.} Correlation functions and pion form factors of local gauge-invariant operators are thus the natural bridge between the bootstrap, which lives in the world of hadrons, and the microscopic QCD Lagrangian, which lives in the world of quarks and gluons.

\subsection*{A mixed system of local probes}

We initiate this program in the simplest possible setting. We consider large $N$ QCD in the chiral limit and probe it with the conserved vector current $J_V^\mu$ of the unbroken $U(N_f)_V$ flavor symmetry. The current is a privileged probe: its dimension is protected along the RG flow, so it retains a sharp meaning in the low-energy effective theory of pions, where its matrix elements are computed by coupling the chiral Lagrangian to background gauge fields. Our bootstrap system comprises three observables: the current two-point function $\Pi(s)$ (the hadronic vacuum polarization), the pion vector form factor $F(s)$, and the $2\to 2$ pion amplitude $M(s,u)$ studied in \cite{Albert:2022oes}.

At large $N$, all three observables are meromorphic, with poles dictated by the exchange of single mesons; the spectral densities $\rho^{(\gamma\gamma)}$, $\rho^{(\gamma 2\pi)}$ and $\rho^{(4\pi)}_J$ reduce to sums of delta functions supported on the meson masses. While the form factor spectral density has no definite sign, unitarity organizes the three densities into a $2\times 2$ positive semidefinite matrix in the spin-one channel, in a large $N$ incarnation of the framework of \cite{Karateev:2019ymz}. Dispersion relations then convert the low-energy data -- the coefficients $a_k$ of $\Pi(s)$, the coefficients $b_k$ of $F(s)$ (with $b_0=1$ fixed by symmetry and $b_1$ proportional to the pion charge radius), and the chiral Lagrangian couplings $g_{n,\ell}$ -- into positive high-energy averages, supplemented by the null constraints encoding crossing symmetry of the amplitude \cite{Tolley:2020gtv,Caron-Huot:2020cmc}. The stage is then set for semidefinite programming.

We obtain bounds of two conceptually distinct kinds.
The first kind is completely universal, and rigorous within the standard large $N$ axioms. The only dynamical input is the high-energy \emph{scaling} of the observables, which fixes the number of subtractions: the logarithmic growth of $\Pi(s)$ dictated by the identity term in the OPE, the Brodsky--Farrar decay $F(s) = O(s^{-1})$, and the usual Regge boundedness of $M(s,u)$. Nothing else about asymptotic freedom is used -- in particular, not the value of any high-energy coefficient. These assumptions allow us to bound ratios of low-energy coefficients, such as the normalized form factor couplings $\tilde b_k \equiv b_k M^{2(k-1)}/\sqrt{a_1 g_{1,0}}$, and to carve out exclusion plots whose geometry we understand analytically in terms of simple (unphysical) solutions to crossing. Refined spectral assumptions -- a rho meson, and then an $f_2$ above it -- sharpen the bounds and reveal imprints of the extremal solutions found in \cite{Albert:2022oes,Albert:2023seb}.

The second kind of bounds inputs asymptotic freedom quantitatively, and is necessarily more phenomenological in character. The idea has a venerable history, dating back to the QCD sum rules of Shifman, Vainshtein and Zakharov (SVZ) \cite{Shifman:1978bx,Shifman:1978by}. One assumes that there is a finite scale $\Lambda$ at which the perturbative computation of $\Pi(s)$ (together with the first power corrections from vacuum condensates) already approximates the exact answer well on the circle $|s|\sim\Lambda^2$ in the complex plane. Dispersive contour integrals can then be \emph{evaluated} at that scale, rather than pushed to infinity, yielding relations between the perturbative UV and the hadronic IR. In the traditional implementation, one further commits to an ansatz for the spectral density -- typically the lowest resonance plus a perturbative continuum above $\Lambda\sim 1\,$--$\,1.5$ GeV -- and extracts estimates for resonance masses and couplings. These estimates turn out to be remarkably accurate, arguably more accurate than they have any right to be. In retrospect, their success can be traced to a happy experimental fact: the measured cross section $\sigma(e^+e^-\to\text{hadrons})$, which is proportional to $\rho^{(\gamma\gamma)}$, shows sharp low-lying resonances followed by a surprisingly early onset of the perturbative plateau (see e.g.~\cite{Davier:2019can}). We include a short review of the traditional SVZ approach in appendix \ref{appendix:svz}, and refer to \cite{Reinders:1984sr,Colangelo:2000dp,Gubler_2019} for a modern perspective.\footnote{For a recent application of two-point sum rules to the glueball sector of three-dimensional Yang--Mills theory, with a careful discussion of their non-rigorous character, see \cite{Caron-Huot:2023tpw}.}

Our SVZ-like analysis upgrades this logic into a bootstrap framework. We make no ansatz whatsoever for the spectral densities: between the cutoff $M=m_\rho$ and the perturbative scale $\Lambda$ they are constrained only by unitarity, in the matrix sense described above.
In this first iteration of the program, we only consider the leading UV behaviors: the one-loop logarithmic growth of $\Pi(s)$, and the vanishing of $F(s)$. 
The price of keeping the UV arcs at finite $\Lambda$ is that the resulting sum rules carry a (systematically-improvable) $\Lambda$-dependent error, controlled by perturbative corrections of order $1/\log\Lambda^2$, as well as non-perturbative corrections from the condensates in $\Pi(s)$ and the pion distribution amplitudes in $F(s)$.
Moreover, rather than committing to a particular value of $\Lambda$, as is customary in SVZ-type analyses, we find it more honest to report every bound as a function of $\Lambda$.
The reader may then decide at which scale to start trusting one-loop perturbation theory, and read off the corresponding bound.

The strategy of marrying perturbative QCD with the form factor bootstrap was first advocated and developed by He and Kruczenski \cite{He:2023lyy,He:2024nwd,He:2025gws} (inspired by \cite{Karateev:2019ymz,Chen:2021bmm,Chen:2021pgx,Correia:2022dyp,Cordova:2023wjp}), in the framework of the non-perturbative S-matrix bootstrap at finite $N$ \cite{Paulos:2016fap,Paulos:2016but,Paulos:2017fhb,Homrich:2019cbt,Guerrieri:2018uew,Guerrieri:2020bto}, with the ambitious goal of (approximately) predicting real-world pion dynamics from QCD inputs.\footnote{See \cite{Guerrieri:2024jkn} for another recent phenomenological application of S-matrix bootstrap methods.} Our philosophy is different. We work strictly at large $N$, where meromorphy renders the analytic structure rigorous and the bootstrap problem sharply posed, and even in our SVZ-type analysis our primary interest is in the conceptual framework -- which assumptions purchase which bounds, with errors under parametric control -- rather than in inputting detailed phenomenology. We view the two approaches as complementary.

Let us now give a brief overview of our main results, to be developed at length in the main text.

\subsection*{Summary of results}

On the rigorous side, the mixed system gives access to the low-energy coefficients of $\Pi(s)$ and $F(s)$, which were invisible to the pure amplitude bootstrap. We find the simple analytic bounds $0\leq a_k M^{2(k-j)} \leq a_j$ for $j \leq k$ and $|\tilde b_1| \leq \sqrt 2$, as well as the numerical bound $|\tilde b_k| \leq 1.246\ldots$ for all $k\geq 2$. The corresponding exclusion plot in the $(\tilde b_1,\tilde b_2)$ plane (figure~\ref{fig:exclusionplot}) is a convex polygon, all of whose corners we analytically understand: they are saturated by the same unphysical amplitudes (the $su$-pole amplitude and its relatives) that already played a starring role in \cite{Albert:2022oes}. Imposing the existence of the rho, and then of the $f_2$ at its real-world mass ratio, sharpens the picture: the bound on $\tilde b_2$ as a function of the gap after the $f_2$ drops sharply at the location of the $f_2$ kink of \cite{Albert:2023seb} (figure \ref{fig:b2vsMtilde-withf2}), showing that this distinguished extremal solution leaves its imprint on the form factor observables as well. While these bounds are new, they do not uncover new extremal theories; this was to be expected, as $\Pi(s)$ and $F(s)$ contribute no null constraints of their own. The true added value of the local probes lies elsewhere, in their calculable ultraviolet.

The SVZ-type sum rules make good on that promise. The crucial novelty is the antisubtracted sum rule for $a_{-1} = \frac{N}{24\pi^2}\Lambda^2$, fully determined by the UV contribution from leading-order perturbation theory.
Normalizing the symmetry-protected coefficient $b_0=1$ by $\sqrt{a_{-1}\, g_{1,0}}$ then yields a bound on a combination of \emph{known} quantities: an upper bound on $f_\pi/(m_\rho \sqrt N)$ as a function of $\Lambda/m_\rho$ (figure \ref{fig:boundonfpi}). We find this result conceptually striking: $g_{1,0}=1/(2f_\pi^2)$ is a ``dimension-six'' coupling, structurally inaccessible to standard positivity bounds, which only constrain ratios starting at ``dimension eight''. The form factor system, supplemented with one-loop asymptotic freedom, bounds the pion decay constant itself, in units of the rho mass.

This bound can be confronted with data. Recent lattice simulations based on the twisted Eguchi--Kawai reduction \cite{Bonanno:2025hzr} determine $f_\pi/(m_\rho\sqrt N) \simeq 0.0713(34)$ in the chiral and continuum limit, at values of $N$ as large as $841$. The lattice value is allowed for large $\Lambda$ but excluded as $\Lambda$ decreases, which we read as a lower bound on the scale at which one-loop perturbation theory may set in: $\Lambda \gtrsim 1.24\, m_\rho$, i.e.\ $\Lambda \gtrsim 960$ MeV using the real-world rho mass for illustration. Assuming the asymptotic freedom result below this scale is inconsistent with unitarity, crossing, and the lattice input -- it would rule out QCD. It is amusing that the bootstrap stops just short of excluding the traditional SVZ window $\Lambda\sim 1\,$--$\,1.5$ GeV. Repeating the analysis at two loops, with the lattice determination of the 't Hooft coupling fed in, shifts the bound only mildly (to $\Lambda \gtrsim 900$ MeV), giving us confidence that the one-loop bounds are meaningful.

Two further applications follow the same template. First, normalizing the pion charge radius $b_1 = \frac{1}{6}\langle r_\pi^2\rangle$ by the SVZ-determined $a_{-1}$, and inputting the lattice value of $f_\pi/\sqrt N$, we obtain an upper bound on $\langle r_\pi^2\rangle m_\rho^2$ as a function of $\Lambda$ (figure \ref{fig:pion-radius-bound}). At large $N$ this is a bound on the chiral Lagrangian coefficient $\kappa_3$ (the large $N$ avatar of $L_9$), which was inaccessible both to pion scattering \cite{Albert:2022oes} and to the pion-photon mixed system \cite{Albert:2023jtd}. Second, and most interestingly, we study how the UV input \emph{backreacts} on the original exclusion plots for the four-derivative pion couplings. Fixing $b_0/\sqrt{a_{-1}\, g_{1,0}}$ to its lattice value forces the form factor system to couple to the amplitude, and the allowed region in the $(\tilde g_{2,0}, \tilde g_{2,1})$ plane shrinks as $\Lambda$ is lowered (figure \ref{fig:shrunk-eft}). Remarkably, it migrates precisely towards the region singled out in \cite{Albert:2023seb} by fixing the low-lying spectrum (rho and $f_2$) to its real-world values -- with no UV input whatsoever. There was no a priori reason for these two very different inputs, one ultraviolet and one infrared, to select overlapping regions. We take their agreement as nontrivial evidence that the asymptotic freedom of QCD is intricately woven into the low-energy data of the pion Lagrangian.

\bigskip

\noindent
The remainder of the paper is organized as follows. In section \ref{sec:Setup} we define the three observables, derive the large $N$ unitarity constraints in matrix form, and discuss their low- and high-energy behavior, including a review of the perturbative control of $\Pi(s)$ and $F(s)$. In section \ref{sec:rigorousBounds} we derive the universal bounds and the associated exclusion plots, with and without spectral assumptions. In section \ref{sec:PhenoBounds} we set up the SVZ-like sum rules and derive the phenomenological bounds on $f_\pi/\sqrt N$, on the perturbative scale, on the pion charge radius, and on the space of chiral Lagrangians. We conclude in section \ref{sec:conclusions} with a discussion of future directions. Appendix \ref{appendix:svz} contains a brief review of the traditional SVZ approach.

\section{Setup}
\label{sec:Setup}
We are concerned with the same theory studied in \cite{Albert:2022oes}, namely four-dimensional large-$N$ QCD in the chiral limit of $N_f$ massless quarks. This theory undergoes spontaneous chiral symmetry breaking with pattern \cite{Coleman:1980mx}
\begin{equation}\label{eq:symm-breaking}
    U(N_f)_L\times U(N_f)_R \longrightarrow U(N_f)_\text{diag} \equiv U(N_f)_V\,,
\end{equation}
resulting in $N_f^2$ Goldstone bosons $\pi^a$ --the massless ``pions''. These are the lightest states in the infinite collection of mesons composing the spectrum of the theory. Beyond scattering amplitudes of these mesons, one can exploit the unbroken symmetry and consider correlation functions of the vector current $J_{Va}^\mu$. See \cite{Albert:2023jtd} for the study of the four-point functions $\langle J_VJ_VJ_VJ_V\rangle$, $\langle J_VJ_V|\pi\pi\rangle$, $\langle J_V|\pi\pi\pi\rangle$ of a $U(1)_V$ subgroup, presented as a mixed scattering system of pions and probe \textit{on-shell} photons. In this paper, we will consider much simpler --but ever so interesting-- observables: the full $U(N_f)_V$ \textit{off-shell} two-point function $\langle J_{V} J_{V}\rangle$ and form factor $\langle J_{V}|\pi\pi\rangle$, combined with the familiar pion four-point amplitude $\langle \pi \pi| \pi \pi\rangle$.

\subsection{Definitions}
We start by defining each of these observables and discussing their main features.

\subsubsection{Two-point function}
The first observable that we will consider is the Fourier transform of the time-ordered two-point function of vector currents,
\begin{equation}\label{eq:2ptFourier}
    -i\Delta_{ab}^{\mu\nu}(p)\equiv \int d^4 x\, e^{-ip\cdot (x-y)}\langle 0|T\{J_{Va}^{\mu}(x)J_{Vb}^{\nu}(y)\}|0\rangle\,.
\end{equation}
This can be regarded as the full non-perturbative propagator of an off-shell probe ``photon'' (or rather its non-Abelian counterpart, a $U(N_f)_V$ ``gluon''), and it often goes by the name of \textit{hadronic vacuum polarization}. As usual, such integrals ought to be defined from analytic continuations of absolutely convergent integrals, but we will not dwell on these (standard) details here. The dependence on the Lorentz indices $\mu,\nu$ is fixed by the Ward identity $p_\mu \Delta_{ab}^{\mu\nu}(p) = 0$, and the dependence on the flavor indices is trivial because there exists only one $U(N_f)$ invariant tensor with two adjoint indices. It is therefore sufficient to consider a scalar function $\Pi(s)$ related to the propagator by
\begin{equation}\label{eq:PiDef}
    \Delta_{ab}^{\mu\nu}(p) \equiv (p^\mu p^\nu - p^2 \eta^{\mu\nu} )\delta_{ab}\Pi(-p^2)\,.
\end{equation}

The analytic structure of $\Pi(s)$ in the complex $s$ plane is under control. It follows from standard arguments leveraging causality that $\Pi(s)$ is, in general, \textit{analytic} everywhere in the complex $s$ plane away from the positive real $s$ axis (see e.g.~\cite{Weinberg:1995mt}). There, the function has a cut with discontinuity proportional to the $J_V J_V$ spectral density,\footnote{The imaginary part is equivalent to the discontinuity because the function is moreover real-analytic, i.e.\ $\Pi(s)^* = \Pi(s^*)$.}
\begin{equation}
    \text{Im}\, \Pi(s) \equiv \frac{1}{2i}\left(\Pi(s+i\epsilon) -\Pi(s-i\epsilon)\right) = \pi \rho^{(\gamma\gamma)}(s) \quad \text{for } s\in \mathds R,\, s>0\,.
\end{equation}
The spectral density is defined here as
\begin{equation}\label{eq:rho-def}
    \frac{1}{(2\pi)^3}\theta(p^0) \rho^{(\gamma\gamma)}(-p^2) \equiv \sum_n \delta^{(4)}(p-p_n)\frac{1}{3(-p_n^2) N_f^2}|\langle0|J_{Va}^\mu(0)|n\rangle|^2\,,
\end{equation}
and it counts physical states that couple to $J_V$, weighted by their overlap with the current.

\begin{figure}[htb]
\centering
\includegraphics[scale=0.45]{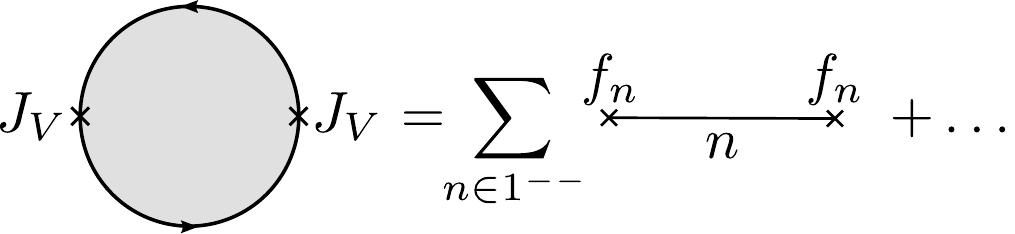}
\caption{Schematic representation of the large $N$ current two-point function in terms of (lhs) planar diagrams and (rhs) single meson exchanges. The dots denote contact terms, which only give analytic contributions to $\Pi(s)$. }
\label{fig:JJ}
\end{figure}

At large $N$, the analytic structure of $\Pi(s)$ simplifies even further: it becomes \textit{meromorphic}, with poles associated to single-meson exchanges. This follows, as usual, from 't~Hooft's large $N$ counting of diagrams \cite{tHooft:1973alw}, through an argument completely parallel to that of \cite{Witten:1979kh} for scattering amplitudes. Since $J_V$ corresponds to a quark bilinear, the only diagrams that survive in the planar limit have the topology of a disk with a quark line running along the boundary, where the two currents are attached (see figure \ref{fig:JJ}). This is the limit that we will be working in. By cutting any of these diagrams to analyze the intermediate states, we discover states with only one quark-antiquark pair, which we identify with \textit{single meson states}. Angular momentum, parity and charge-conjugation conservation further fix the quantum numbers of these mesons to $J^{PC}=1^{--}$.\footnote{This is a consequence of the symmetry properties of $J_V$. Since the vector symmetry is not spontaneously broken, $J_V$ does not create massless (Goldstone) particles when acted on the vacuum. From \eqref{eq:PiDef} it is then clear that a massive pole in $\Pi(s)$ results in the massive spin-one propagator,
$$\Delta^{\mu\nu}_{ab}(p) \propto \delta_{ab}\frac{\eta^{\mu\nu}+\frac{p^\mu p^\nu}{m^2}}{p^2+m^2}\,,$$
fixing $J=1$. Under parity, we have $P:J_{Va}^\mu \to +J_{Va}^\mu$, which for a vector is conventionally denoted by $P=-$. Under charge conjugation, $C:J_{Va}^\mu \to -(J_{Va}^\mu)^T$, and so $C=-$.}
This collection of mesons starts with the rho (the lightest state in the leading Regge trajectory), and continues with a whole family of spin-one mesons from subleading trajectories.\footnote{In the literature, various names are used for $1^{--}$ real-world mesons: $\rho_1$, $\omega_1$ or $\phi_1$ depending on their isospin projections. All these projections are degenerate at large $N$.}

We parametrize the probability amplitude of creating such a meson state $|n_a^\lambda(p)\rangle$ from acting the vector current on the vacuum by a coupling $f_n$, which can be taken to be real. 
Explicitly, in terms of the usual massive spin-one polarization vectors, we set
\begin{equation}
    \langle 0|J_{Va}^\mu(0)|n^\lambda_b(p)\rangle \equiv f_n m_n \epsilon^\mu_\lambda (p)\delta_{ab}\,.
\end{equation}
It then follows that, at large $N$, the spectral density \eqref{eq:rho-def} reduces to a collection of delta functions,
\begin{equation}\label{eq:rhogg-deltas}
    \rho^{(\gamma\gamma)}(s) = \sum_{n\in 1^{--}} f_n^2\, \delta(s-m_n^2)\,.
\end{equation}
Note that our conventions are such that $\Pi(s)$ is dimensionless and $f_n$ has dimensions of energy. Moreover, since all planar diagrams with a single quark loop scale linearly with $N$, we have $\Pi(s)\sim N$, and $f_n \sim \sqrt{N}$.

\subsubsection{Form factor}
Next, we proceed to the form factor. In terms of asymptotic pion states, it is defined by
\begin{equation}\label{eq:F-def}
    \mathcal F^\mu_{abc}(p_1,p_2) \equiv \langle 0| J_{Vc}^\mu(0)|\pi_a(p_1)\pi_b(p_2)\rangle\,.
\end{equation}
It represents an off-shell probe ``vector gluon'' decaying into two pions (or, by crossing, the probe gluon striking an otherwise freely moving pion). Alternatively, it can be defined through an LSZ prescription using the fact that the axial current $J_A$ creates (among other states) a pion when acted on the vacuum. This is because $J_A$ is a generator of the symmetry spontaneously broken by \eqref{eq:symm-breaking}, and the pion its associated Goldstone boson. In detail,
\begin{equation}\label{eq:JApi}
    \langle 0|J_{Aa}^\mu(0)|\pi_b(p)\rangle \equiv i f_\pi p^\mu \delta_{ab}\,.
\end{equation}
The form factor can thus be obtained from the three-point function $\langle J_VJ_A J_A\rangle$ by picking out the pion poles emanating from the axial current insertions. Namely,
\begin{equation}\label{eq:FfromLSZ}
    \mathcal F_{abc}^\mu(p_1,p_2) = -\lim_{p_1^2,p_2^2\to 0} \frac{p_{1\nu}}{f_\pi} \frac{p_{2\rho}}{f_\pi}\int d^4 x_1 d^4 x_2\,  e^{-i(p_1\cdot x_1+p_2\cdot x_2)} \langle 0 | T\{J_{Aa}^\nu(x_1)J_{Ab}^\rho(x_2)J_{V c}^{\mu}(0)\} | 0 \rangle\,.
\end{equation}

Either way, Lorentz invariance and the Ward identity $p_\mu \mathcal F^\mu_{abc}=0$ fix the space-time index structure to be proportional to $(p_1-p_2)^\mu$. For the flavor index dependence, there are a priori two possible invariant structures,\footnote{Here $T_a$ are the $\mathfrak u(N_f)$ generators in the defining representation. They include the fundamental $\mathfrak{su}(N_f)$ generators, and the identity normalized by $1/\sqrt{2N_f}$. We use conventions in which $\text{Tr}(T_aT_b)=\frac{1}{2}\delta_{ab}$, and we lower and raise adjoint indices with this Killing metric.}
\begin{equation}
    d_{abc} \equiv 2\text{Tr}(T_a\{T_b,T_c\})\,, \quad \text{or}\quad f_{abc}\equiv \frac{2}{i}\text{Tr}(T_a[T_b,T_c])\,.
\end{equation}
Since $J_V$ has $C=-$, charge conjugation symmetry (which acts on the flavor generators as $T_a\to T_a^T$) picks out $f_{abc}$. We can thus write\footnote{The factor of $i$ was introduced so that $F(s)$ is real-analytic, i.e.\ $F(s)^* = F(s^*)$. This follows from rewriting the form factor as
$$\mathcal F^\mu_{abc}(p_1,p_2) \equiv \langle \pi_b(-p_2)| J_{Vc}^\mu(0)|\pi_a(p_1)\rangle\,,$$
where we traded an incoming pion for an outgoing one with opposite momentum, and then noting that $\mathcal F_{abc}^{\mu}(p_1,p_2)^* = \mathcal F_{bac}^{\mu}(-p_2^*,-p_1^*)$.
}
\begin{equation}\label{eq:F(s)-def}
    \mathcal F^\mu_{abc}(p_1,p_2) \equiv i(p_1-p_2)^\mu f_{abc} F(-(p_1+p_2)^2)\,.
\end{equation}

\begin{figure}[htb]
    \centering
    \includegraphics[scale=0.45]{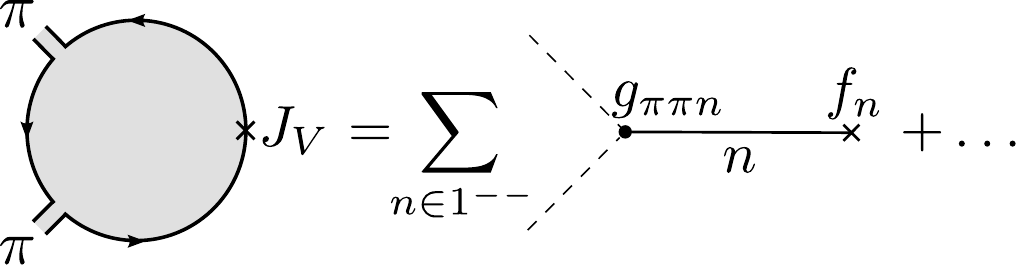}
   \caption{Schematic representation of the large $N$ form factor in terms of (lhs) planar diagrams and (rhs) single meson exchanges. The dots denote contact terms, which only give analytic contributions to $F(s)$.}
\label{fig:2piJ}
\end{figure}

Much like $\Pi(s)$, $F(s)$ is in general analytic in the complex $s$ plane with a cut on the positive real $s$ axis. At large $N$, only disk diagrams contribute, and this function also becomes meromorphic with poles corresponding to $1^{--}$ meson exchanges (see Figure \ref{fig:2piJ}). The corresponding spectral density then reads
\begin{equation}
    \text{Im} F(s) \equiv \pi \rho^{(\gamma2\pi)}(s) \propto \sum_{n\in 1^{--}} f_n g_{\pi\pi n} \delta(s-m_n^2)\,,
\end{equation}
where $g_{\pi\pi n}$ is the $\pi\pi\to n$ on-shell coupling. The standard large-$N$ counting rules determine a scaling $g_{\pi\pi n}\sim 1/\sqrt{N}$ \cite{Witten:1979kh}, from which follows that $F(s)\sim N^0$.

\subsubsection{Scattering amplitude}
\begin{figure}[htb]
    \centering
    \includegraphics[scale=0.45]{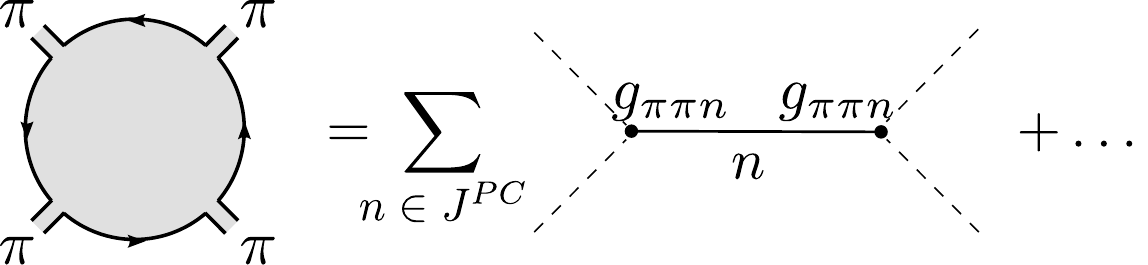}
   \caption{Schematic representation of the large $N$ pion amplitude in terms of (lhs) planar diagrams and (rhs) single meson exchanges. The dots denote contact terms and (some) $u$-channel exchanges. }
\label{fig:amplitude}
\end{figure}

The last object that we will consider is the $2\to 2$ pion scattering amplitude $\mathcal T_{abcd}$, whose properties were thoroughly spelled out in \cite{Albert:2022oes}. At large $N$, this amplitude is controlled by a single $s\leftrightarrow u$ crossing-symmetric function $M(s,u)$ of the Mandelstam invariants. It is meromorphic in the complex $s$ and $u$ planes, with poles corresponding to physical $s$- and $u$-channel meson exchanges (see figure \ref{fig:amplitude}). It has no $t$-channel poles due to the OZI rule \cite{Okubo:1963fa,Zweig:1964jf,Iizuka:1966fk}. Discrete symmetries impose selection rules which force the exchanged mesons to have quantum numbers $J^{PC}=\text{even}^{++}, \text{odd}^{--}$. At fixed $u$, the discontinuity in $s$ on the positive real axis can be expanded in a partial wave expansion,
\begin{equation}\label{eq:PWexp}
    \text{Im}\, M(s,u) = \sum_J n_J \rho_J^{(4\pi)}(s)\mathcal P_J\left(1+\frac{2u}{s}\right)\,,
\end{equation}
where $\mathcal P_J(x)$ are Legendre polynomials and $n_J\equiv 16\pi (2J+1)$. The meromorphicity of $M(s,u)$ implies that the spectral density once again takes the form of a sum of delta functions,
\begin{equation}
    \rho_J^{(4\pi)}(s) \propto \sum_{n\in\left\{\begin{smallmatrix}
        \text{even}^{++}\\
        \text{odd}^{--}
    \end{smallmatrix}\right.\hspace{-30pt}}
    g_{\pi\pi n}^2 \delta(s-m_n^2)\,.
\end{equation}
From the scaling of $g_{\pi\pi n}$, it is clear that the amplitude scales as $M(s,u)\sim 1/N$, as befits disk diagrams with four boundary external legs.

\subsection{Unitarity}\label{sec:Unitarity}
The key assumption that will drive the bootstrap is that the underlying theory is \textit{unitary}. It is well known that unitarity enforces positivity for the spectral density of the $2\to 2$ amplitude, i.e.\ $\rho^{(4\pi)}_J(s)\geq 0$ for $s\geq 0$. The same is true for the two-point function spectral density $\rho^{(\gamma\gamma)}(s)$, as follows from the definition \eqref{eq:rho-def}. But this ceases to be true for the form factor, essentially because the initial and final states in \eqref{eq:F-def} are different. A strategy to extend unitarity constraints to form factors was introduced in \cite{Karateev:2019ymz}. The idea is to construct a matrix of inner products of the state $|J_V\rangle$ and the asymptotic states $|\pi\pi\rangle_\text{in}$, $|\pi\pi\rangle_\text{out}$, which must be positive semidefinite in a unitary theory. In this way, the two-point function and the $S$-matrix come in to ``complete the square'' and impose unitarity constraints on $F(s)$.

The constraints that follow from this procedure take the schematic form \cite{Karateev:2019ymz}
\begin{equation}
\begin{gathered}
    |S_J(s)|^2\leq 1\,, \qquad \rho^{(\gamma\gamma)}(s) \geq |F(s)|^2\,, \\
    \rho^{(\gamma\gamma)}(s)(1-|S_1(s)|^2)-2|F(s)|^2+F^*(s)^{2}S_1(s) + F(s)^2S_1^*(s)\geq 0\,,
\end{gathered}    
\end{equation}
where we have absorbed some normalization factors by slight redefinitions of the functions. At large $N$, most of these constraints trivialize. Indeed, writing the partial $S$-matrix in terms of the partial wave, $S_J(s)=1+ia_J(s)$, and using that $a_J(s)\sim 1/N$, we see that the first constraint reduces to $\text{Im}\, a_J(s) \propto \rho^{(4\pi)}_J(s)\geq 0$. In turn, since $\rho^{(\gamma\gamma)}(s)\sim N$ but $F(s)\sim N^0$, the second constraint is trivially satisfied. From the last constraint, we get $\rho^{(\gamma\gamma)}(s)\rho^{(4\pi)}_1(s)\gtrsim \rho^{(\gamma2\pi)}(s)^2$ (again up to normalization factors), which provides the only constraint on the large $N$ form factor spectral density.

Let us now carefully fix the relative normalization constant in this inequality, which is meaningful because it is invariant under rescalings of the pion wavefunction and the conserved current. This is simplest by computing the precise contribution of a single $1^{--}$ meson $n_\mu^c$ to each observable. The relevant interaction terms are
\begin{equation}\label{eq:Lint}
    \mathcal L_{\text{int}} \supset -g_{\pi\pi n} f_{abc} \pi^a\partial^\mu \pi^b n_\mu^c +  f_n m_n e A_{V\mu}^a n^\mu_a\,,
\end{equation}
where $A_{V\mu}^a$ is a background gauge field for the unbroken $U(N_f)_V$ symmetry. Since the gauge field couples to the conserved current by a term $\int d^{4}x\, e A_{V\mu}^a(x) J^\mu_{Va}(x)$ in the action $S[A_V]$, we can compute correlation functions of currents as
\begin{equation}\label{eq:currents-gauge}
    \Big< J^\mu_{Va}(x)\cdots \Big>_S = \frac{1}{e}\frac{\delta}{i\delta A_\mu^a(x)} \Big< \cdots \Big>_{S[A_V]}\Big|_{A_V=0}\,.
\end{equation}
To compute correlation functions of $J_V$ (in momentum space), we thus just need to compute Feynman diagrams where each current is replaced by a vector gluon, and divide by a factor of $i e$ per leg.

Following this prescription, it is straightforward to compute
\begin{subequations}\label{eq:spin1exchange}
\begin{align}
    -i\Delta^{\mu\nu}_{ab}(p) =& -i \left(p^\mu p^\nu + m_n^2 \eta^{\mu\nu}\right)\delta_{ab} \frac{f_n^2}{m_n^2 - s}\,,\\
    \mathcal F_{abc}^\mu(p_1,p_2) =&\, i(p_1 -p_2)^\mu f_{abc} \frac{g_{\pi\pi n}f_n m_n}{m_n^2 - s}\,.
\end{align}
Similarly, the $2\to 2$ amplitude is readily evaluated to \cite{Albert:2022oes}
\begin{equation}\label{eq:spin1amplitude}
    M(s,u) = \frac{1}{2} g_{\pi\pi n}^2 m_n^2 \left(\frac{\mathcal P_1\left(1+\frac{2u}{m_n^2}\right)}{m_n^2-s} + \frac{\mathcal P_1\left(1+\frac{2s}{m_n^2}\right)}{m_n^2-u}\right)\,.
\end{equation}
\end{subequations}
Comparing these expressions to the general forms \eqref{eq:PiDef}, \eqref{eq:F(s)-def}, \eqref{eq:PWexp}, and extracting the imaginary parts, we determine
\begin{subequations}\label{eq:1--rhos}
\begin{align}
    \rho^{(\gamma\gamma)}(s) =&\, f_n^2  \delta(s-m_n^2)\,,\\
    \rho^{(\gamma2\pi)}(s) =&\, g_{\pi\pi n}f_n m_n\delta(s-m_n^2)\,,\\
    n_1 \rho_1^{(4\pi)}(s) = &\, \frac{\pi}{2}g_{\pi\pi n}^2 m_n^2 \delta(s-m_n^2)\,.
\end{align}
\end{subequations}
This is the contribution of a generic $1^{--}$ meson to each spectral density. The full densities are the sum of the contributions of all such mesons.

The statement of unitarity at the level of the Lagrangian \eqref{eq:Lint} is that it be Hermitian. This requires the coefficients $g_{\pi\pi n}, f_n$ to be real. It then immediately follows from \eqref{eq:1--rhos} that the spectral densities satisfy the inequality
\begin{equation}
    \rho^{(\gamma\gamma)}(s) \frac{n_1}{\pi}\rho_1^{(4\pi)}(s) \geq \frac{1}{2} \rho^{(\gamma2\pi)}(s)^2\,,
\end{equation}
or, equivalently, that the matrix of spectral densities is positive-semidefinite,
\begin{equation}\label{eq:rho1}
    \tilde \rho_1(s)\equiv \begin{pmatrix}
        \rho^{(\gamma\gamma)}(s) & \frac{1}{\sqrt{2}}\rho^{(\gamma 2\pi)}(s)\\
        \frac{1}{\sqrt{2}}\rho^{(\gamma 2\pi)}(s) & \frac{n_1}{\pi}\rho_1^{(4\pi)}(s)
    \end{pmatrix}\succeq 0\,.
\end{equation}

\subsection{Low-energy behavior}\label{sec:LEbehavior}
At low energies, specifically below the mass $M$ of the first exchanged meson in the spectrum (the rho), all relevant correlators are analytic functions of the Mandelstam invariants, as no physical intermediate state can go on-shell. In this regime, $s<M^2$, the two-point function and the form factor can be expressed as polynomials in $s$, 
\begin{equation}\label{eq:lowEexp}
    \Pi_{\text{low}}(s)= a_0 + a_1 s + a_2 s^2 +\dots \, , \quad 
    F_\text{low}(s) = b_0 + b_1 s+ b_2 s^2 + \dots \,. 
\end{equation}
In our conventions, both $\Pi(s)$ and $F(s)$ are dimensionless, so the effective field theory coefficients $a_k$, $b_k$ carry mass dimension $-2k$. Similarly, the low-energy scattering amplitude $M_{\text{low}}(s,u)$ is a symmetric polynomial in $s$ and $u$, with an expansion \cite{Albert:2022oes}
\begin{equation}\label{eq:pion-eft}
    M_{\text{low}}(s,u) = g_{1,0}(s+u) + g_{2,0}(s^2+u^2) + 2g_{2,1}su + \dots \,,
\end{equation}
around $s,u \sim 0$. The coefficients $g_{k,l}$ have mass dimension $-2k$.

The coefficients in \eqref{eq:pion-eft} correspond to effective four-point contact interactions generated by integrating out the massive mesons in the spectrum. Such interactions furnish the EFT of massless pions, and are compactly collected in the celebrated chiral Lagrangian \cite{Gasser:1983yg,Gasser:1984gg,Kaiser:2000gs}. Similarly, the coefficients in \eqref{eq:lowEexp} capture the two-point function and form factor of $J_V$ in the low-energy EFT. It is meaningful to talk about these because conserved currents are protected under RG flows. They are obtained by coupling the effective Lagrangian to the corresponding background gauge fields. The chiral Lagrangian with background gauge fields $A_\mu^L$ and $A_\mu^R$ associated to the full chiral symmetry $U(N_f)_L \times U(N_f)_R$ was reviewed e.g.\ in \cite{Donoghue:1992dd,Albert:2023jtd}. For our purposes, we restrict to the non-abelian vector background, obtained by setting
\begin{equation}
  A_\mu^L =  A_\mu^R = eA_\mu^a T^a\,.
\end{equation}
With this choice, the chiral Lagrangian becomes
\begin{align}\label{eq:Lch}
S_{\mathrm{Ch}}[A]=\int d^4 x[ & -\frac{f_\pi^2}{4} \tr{D_\mu U D^\mu U^{\dagger}}+\kappa_1 \tr{D_\mu U D^\mu U^{\dagger} D_\nu U D^\nu U^{\dagger}} \\
& +\kappa_2 \tr{D_\mu U D_\nu U^{\dagger} D^\mu U D^\nu U^{\dagger}} \nonumber\\
& +i \kappa_3 \tr{ F_{\mu \nu}\left[D^\mu U, D^\nu U^{\dagger}\right]} \nonumber\\
& \left.+ \kappa_4  \tr{F_{\mu \nu} U F^{\mu \nu}  U^{\dagger}}+ \kappa_5^{\prime} \tr{F_{\mu \nu} F^{\mu \nu}}+\cdots\right], \nonumber
\end{align}
where we introduced the covariant derivative
\begin{equation}
    D_{\mu}U \equiv \partial_\mu U + ie A_\mu^a [T^a, U] \;,
\end{equation}
and the field strength
\begin{equation}
    F^a_{\mu \nu} \equiv e\left(\partial_\mu A^a_\nu - \partial_\nu A^a_\mu \right) -e^2 f^{abc}A_\mu^b A_\nu^c \quad , \quad F_{\mu \nu } = F_{\mu \nu}^a T^a  \;.
\end{equation}
The field $U(x)=\exp{\left(\frac{2i}{f_{\pi}} T^a \pi^a(x)\right)}$ is the unitary matrix which encodes the pion fields $\pi^a$ as the Goldstone bosons of the spontaneous symmetry breaking $U(N_f)_L\times U(N_f)_R \to U(N_f)_V$. At large $N$, only single trace operators are present in the action, and --as in any EFT-- terms are organized in a derivative expansion, with the interactions in each higher-order term suppressed by further powers of the momenta.

The decay constant $f_\pi$ and the couplings $\kappa_1, \kappa_2,\dots$ in the chiral Lagrangian are unknown Wilson coefficients describing the IR tail of an RG flow. They capture (in an indirect way) UV information of the underlying theory, and we will derive bounds for them using $S$-matrix bootstrap methods.
When expanded in pion fields, one can match these coefficients with the analytic expansions in \eqref{eq:lowEexp} and \eqref{eq:pion-eft}. The first terms in the amplitude were matched in \cite{Albert:2022oes,Albert:2023jtd};
\begin{equation}
    g_{1,0}=\frac{1}{2f_\pi^2}, \quad g_{2,0} = \frac{2\kappa_1 +4 \kappa_2}{2 f_\pi^4}, \quad g_{2,1}= \frac{4\kappa_2}{f_\pi^4} \;.
\end{equation}
For the form factor, we need to look for terms with two pions and a gauge field in the chiral Lagrangian. Expanding in pion fields, the relevant interaction terms at the lowest order in derivatives are
\begin{equation}
    e f^{abc}\pi^a(\partial^\mu\pi^b)A_\mu^c -\frac{4e\kappa_3}{f_\pi^2}f^{abc}(\partial^\mu\pi^a)(\partial^\nu \pi^b)(\partial_\mu A_\nu^c) \subset \mathcal{L}_{\text{int}} \;.
\end{equation}
The first term follows from the kinetic term of the pions, and the second term is the $\kappa_3$ interaction. Using the definition in $\eqref{eq:F(s)-def}$ and $\eqref{eq:currents-gauge}$, we find
\begin{equation}\label{eq:Flow}
    F(s) = 1 + \frac{2\kappa_3}{f_\pi^2}s +\dots \;,
\end{equation}
fixing $b_0$ and $b_1$ from \eqref{eq:lowEexp}. The condition that $F(0)=b_0=1$, which will play a role below, can be directly related to the statement that the pion transforms in the adjoint representation of $U(N_f)_V$, see e.g.\ \cite{Weinberg:1995mt}.

Using similar methods, we calculate the propagator for two off-shell vector gluons to~be
\begin{equation}
    \Pi(s)= -2(\kappa_4 + \kappa_5') + \dots\,,
\end{equation}
fixing the order $s^0$ constant, $a_0$, in \eqref{eq:lowEexp}. Note that the term $\kappa_5'$ in \eqref{eq:Lch} is a \textit{counterterm}, i.e.\ it does not depend on any dynamical fields. In general, we are always free to add such a term (and higher-derivative versions thereof) to the action without changing the physics. One might thus worry that the coefficients $a_i$ are not physical and can be tuned at will. This is not the case because counterterms can only be fixed once for any given RG flow. In the following section, we will give the definition of $\Pi(s)$ in terms of UV currents of the QCD Lagrangian. This effectively fixes the counterterms in the UV. When flowing to the IR, $\kappa_5'$ and higher-derivative counterterms are generated, but we no longer have the freedom to remove them, and so they are physical and-- in principle-- measurable.

\subsection{High-energy behavior}\label{sec:HEbehavior}
We finish this section by discussing the limit of $s\to \infty$ of the various functions.

\subsubsection{Two-point function}
The high-energy behavior of $\Pi(s)$ is controlled by the short-distance limit of the two-point function $\langle J_{Va}^\mu(x) J_{Vb}^\nu(0)\rangle$. Two-point functions of local operators $\langle \mathcal O(x)\mathcal O(0)\rangle$ in QFT (rather than CFT) are interesting because they are one-parameter functions which probe the full RG flow. For instance, Zamolodchikov famously used them to prove his $c$-theorem \cite{Zamolodchikov:1986gt} (see also \cite{Cappelli:1990yc,Karateev:2020axc}). Consider first the case of a CFT$_\text{UV}$ with a relevant deformation $g\int d^d x\, \phi(x)$. In the limit of $x\to 0$, the theory nears its UV fixed point, and the two-point function can be expanded in conformal perturbation theory. In particular, the theory ``inherits'' an operator product expansion (OPE) $\mathcal O(x) \mathcal O(0) \sim \sum_k C_k(x) \mathcal O_k(0)$ from the UV CFT \cite{Zamolodchikov:1990bk}. Since expectation values on the IR vacuum may not vanish, however, the two-point function becomes an expansion in one-point functions of the form
\begin{equation}\label{eq:OOOPE}
    \langle \mathcal O(x) \mathcal O(0) \rangle \sim \sum_k C_k(x)\langle \mathcal O_k(0)\rangle\,.
\end{equation}
While the one-point functions are non-perturbative data that need to be fixed e.g.\ with experiments or numerical simulations, the structure functions $C_k(x)$ are under perturbative control. Scaling invariance requires that they take the form
\begin{equation}
    C_k(x) = \frac{c_k(g|x|^{d-\Delta_\phi})}{|x|^{2\Delta_{\mathcal O}-\Delta_k}}\,,
\end{equation}
where $\Delta_{\phi},\Delta_{\mathcal O},\Delta_k$ are respectively the UV scaling dimensions of $\phi,\mathcal O, \mathcal O_k$.
The statement is that the $c_k$ are analytic functions of the coupling $g$, and so can be computed order by order in (conformal) perturbation theory \cite{Zamolodchikov:1990bk}.

In four-dimensional QCD, the situation is somewhat more involved because the gauge coupling is only marginally relevant. Nevertheless, we still expect an OPE which is under perturbative control, owing to the asymptotic freedom of the theory \cite{Gross:1973id,Politzer:1973fx}. While there is no rigorous non-perturbative proof of the OPE in QCD, its validity is well established, and it has had tremendous success since Wilson proposed it in 1969 \cite{Wilson:1969zs} (see e.g.\ \cite{Weinberg:1996kr} for a textbook introduction). Listing the gauge invariant scalar operators of QCD with nonzero condensates, one infers the short-distance behavior \cite{Bernard:1975cd,Shifman:1978bx,Shifman:1978by}
\begin{equation}
    \langle J_{Va}^\mu(x) J_{Vb}^\nu(0)\rangle \sim  \frac{c^{\mu\nu}_{\mathds 1}(\alpha_s,x)}{|x|^{6}}\delta_{ab}  + \frac{c_{\psi \psi}^{\mu \nu}(\alpha_s,x)}{|x|^{3}} \delta_{ab}\langle \bar \psi_i \psi^i \rangle + \frac{c^{\mu\nu}_{GG}(\alpha_s,x)}{|x|^2} \delta_{ab}\langle \text{Tr}(G_{\rho\sigma} G^{\rho\sigma})\rangle + \cdots\,.
\end{equation}
Here $\psi^i$ are the $N_f$ quark fields, $G_{\mu\nu}$ the $SU(N)$ color field strength and $\alpha_s \equiv g_{\text{YM}}^2/4\pi$ the coupling. Performing a Fourier transform and stripping off the tensor structure from \eqref{eq:PiDef}, we find by dimensional analysis the momentum space high-energy behavior,
\begin{equation}\label{eq:pOPE}
    \Pi(s) \sim \tilde c_{\mathds 1}(\alpha_s,s) + \frac{\tilde c_{\psi \psi}(\alpha_s,s)}{s^{3/2}}\langle \bar \psi_i \psi^i \rangle + \frac{\tilde c_{GG}(\alpha_s,s)}{s^{2}} \langle \text{Tr}(G_{\rho\sigma} G^{\rho\sigma})\rangle + \cdots\,.
\end{equation}
The $\tilde c_k(\alpha_s,s)$ can be expanded in integer powers of $\alpha_s$ and are thus perturbative. Note that their $s$-dependence is only through logs because the coupling is classically dimensionless.

\begin{figure}[htb]
\centering
\includegraphics[scale=0.45]{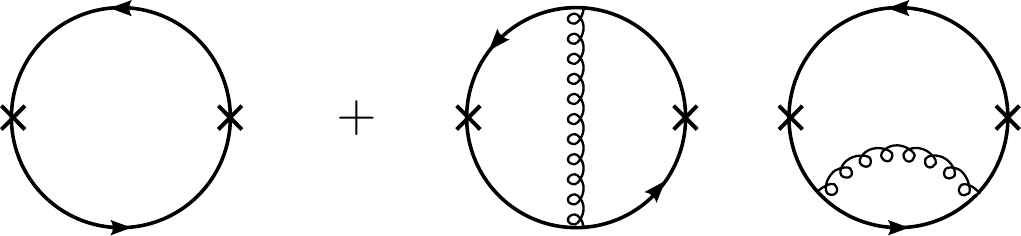}
\caption{One- and two-loop QCD diagrams contributing to $\Pi_{\text{pQCD}}(s)$. Solid lines denote quarks and curly lines denote gluons.}
\label{fig:diags}
\end{figure}

In this paper, we will only be concerned with the identity contribution, which is free from vacuum condensates and thus fully perturbative. $\tilde c_{\mathds 1}(\alpha_s,s)$ can be computed with a standard QCD (planar) loop expansion of the two-point function of vector currents,
\begin{equation}
    J_{Va}^\mu = \bar \psi_j \gamma^\mu (T_a)_i^j \psi^j\,.
\end{equation}
The first diagrams are depicted in figure \ref{fig:diags}. The one-loop result, which amounts to free theory (i.e.\ order $\sim(\alpha_s)^0$), is a textbook computation. It gives
\begin{equation}\label{eq:PipQCD}
    \Pi_{\text{pQCD}}(s)\equiv \tilde c_{\mathds 1}(\alpha_s,s) = -\frac{N}{24\pi^2}\log(-s/\mu^2) + \cdots\,.
\end{equation}
Here, the scale $\mu$ was introduced as a necessary by-product of renormalizing a UV divergence. It came from the finite part of a counterterm $\sim \text{Tr}(F^2)$ of the background gauge fields.\footnote{In position space, it comes from a contact term supported at $x=0$ needed to regulate the diverging Fourier transform. See e.g.\ \cite{Freedman:1991tk} for a particular scheme to cancel such divergences.} We can tune the scale $\mu$, i.e.\ the scheme dependence, by changing this counterterm further. We note that, in writing \eqref{eq:PipQCD}, we have not only fixed this counterterm, but also all of its higher-derivative counterparts, making the IR couplings $a_k$ in \eqref{eq:lowEexp} meaningful.

We will come back to higher-loop corrections in section \ref{sec:HigherLoop}. For now, let us note that, due to running, the gauge coupling drops with energy as $\alpha_s(E)\sim 1/\log\, E^2$. So perturbative corrections to the high-energy behavior of $\Pi(s)$ are logarithmically suppressed. Non-perturbative corrections are due to the condensates in \eqref{eq:pOPE} and are power-law suppressed. In the chiral limit of zero quark mass, the quark condensate structure function $\tilde c_{\psi\psi}$ actually vanishes because it is not compatible with the full chiral symmetry $U(N_f)_L \times U(N_f)_R$.\footnote{The reason is that $\bar\psi \psi$ transforms in a representation which is not contained in the product $J_V\times J_V$. This is most apparent when writing the correlator in terms of $J_L,J_R$ currents.} In consequence, the leading power correction to the two point function is of order $O(1/s^2)$, and it is due to the gluon condensate $\text{Tr} \,G^2$.

We finish by commenting on an apparent puzzle. Namely, we argued that at large $N$ $\Pi(s)$ is a \textit{meromorphic} function with poles on the physical spectrum of spin-one mesons. We also argued that its high-energy behavior is given by \eqref{eq:PipQCD}, but this function has a \textit{cut}. Taking its imaginary part, we derive the asymptotic spectral density
\begin{equation}\label{eq:asymprho}
    \rho^{(\gamma\gamma)}(m^2) \xrightarrow[m^2\to \infty]{} \frac{N}{24\pi^2}\,.
\end{equation}
This was supposed to be a sum over delta functions, recall \eqref{eq:rhogg-deltas}, how should we interpret this continuous result? The situation here is reminiscent of Cardy formula \cite{Cardy:1986ie} in two-dimensional CFT, which also predicts a continuum limit for the asymptotic density of states in theories with a discrete spectrum. The resolution is that these limits apply only on the spectral density after averaging over finite energy windows. The need for this averaging was already anticipated by Poggio, Quinn and Weinberg as early as 1976 \cite{Poggio:1975af}, see \cite{Shifman:2000jv} for a review. Rigorous statements about asymptotic spectral densities should really be framed in terms of Tauberian theorems; see \cite{Pappadopulo:2012jk,Das:2017vej,Mukhametzhanov:2018zja,Mukhametzhanov:2019pzy,Pal:2019zzr,Mukhametzhanov:2020swe,Das:2020uax} for applications in CFT, and in particular \cite{Qiao:2017xif} for a discussion in the context of large $N$ gauge theory.

\subsubsection{Form factor}\label{sec:FormFactorHE}
The high energy behavior of the form factor is also controlled (in part) by asymptotic freedom, albeit in a more intricate way. The subject has a rich history; here we limit ourselves to a brief account of the facts that we will need, see e.g.\ \cite{Sterman:1997sx} for a useful review. The leading decay $F(s)\sim s^{-1}$, compatible with experimental observations, was inferred by Brodsky and Farrar \cite{Brodsky:1973kr} early on using simple counting rules from the number of constituent quarks. This behavior was later put on firmer footing by Efremov, Radyushkin, Brodsky and Lepage (ERBL) \cite{Efremov:1978rn,Efremov:1979qk,Lepage:1979zb,Lepage:1980fj}. They proved, to all orders in perturbation theory, that the leading $s\to \infty$ behavior of the form factor factorizes as
\begin{equation}\label{eq:Ffactorization}
    F(s) = \int_0^1 dx dy\, \phi_\pi(y,\mu^2) T(y,x,s,\mu^2) \phi_\pi(x,\mu^2) + \cdots\,.
\end{equation}
Let us explain the ingredients in this expression. First, the functions $\phi_\pi(x,\mu^2)$ are the so-called leading-twist \textit{pion distribution amplitudes}, defined from the overlap of a pion state with a light-ray operator composed by two null-separated quarks $\bar \psi(0), \psi(x_-)$ with a Wilson line $W[0,x_-]$ stretching between them. Namely,
\begin{equation}\label{eq:light-ray}
    \langle 0| \bar \psi_j(0) W[0,x_-]\gamma_+ \gamma_5(T_a)_i^j\psi^i(x_-)|\pi_b(p)\rangle \propto i f_\pi p_+ \delta_{ab} \int_0^1dy\, e^{-i y p_+ x_- } \phi_\pi(y,\mu^2)\,,
\end{equation}
where the (scheme-dependent) scale $\mu$ arises in renormalizing the light-ray operator, see e.g.\ \cite{Braun:2003rp} for details. In light-cone gauge $A_+=0$, the Wilson line drops, and $\phi_\pi(y,\mu^2)$ can be given the interpretation of the probability amplitude for finding a quark-antiquark pair with momenta $yp, (1-y)p$ inside a pion of momentum $p$. So these functions capture how the UV degrees of freedom of QCD are encoded inside asymptotic meson states, but one should not read them as an \textit{ad hoc} quark model concoction.

Second, $T(y,x,s,\mu^2)$ is a fully perturbative part describing the hard scattering of quarks. At leading order, it is given by a diagram where the probe ``photon'' strikes a pair of quarks exchanging a gluon, which is clearly of order $T\sim \alpha_s(\mu^2)/s$. This already confirms the prediction from counting rules, but it can be refined further using the running of the pion distribution amplitudes. For this, it is convenient to use an expansion in Gegenbauer polynomials,
\begin{equation}\label{eq:moments}
    \phi_\pi(x,\mu^2) = 6x(1-x) \sum_n^\infty \phi_n(\mu^2) C_n^{3/2}(2x-1)\,,
\end{equation}
where $\phi_n(\mu^2)$ are $n$th-moments of $\phi_\pi(x,\mu^2)$. The running of these moments is governed by the so-called ERBL evolution equations \cite{Efremov:1978rn,Efremov:1979qk,Lepage:1979zb,Lepage:1980fj} (see also \cite{Braun:2003rp}), which predict a logarithmic behavior $\phi_n(\mu^2) \sim (\log \mu^2)^{-c_n}$ with known exponents $c_n$. The first moment is special because $c_0=0$, so it is a constant which can actually be fixed by comparing \eqref{eq:light-ray} at $x_-=0$ with \eqref{eq:JApi}. Higher moments have $c_n> 0$, and they depend on a non-perturbative measurement of the moment $\phi_n(\mu_0^2)$ at some reference scale $\mu_0$.

As usual in perturbation theory, the renormalization scale $\mu$ is to be taken around the characteristic scale of the process; $\mu^2 \sim s$ in this case. The high-energy limit of the form factor \eqref{eq:Ffactorization} is thus given by $F\sim s^{-1}$ with the expansion in moments \eqref{eq:moments} on top of it decaying with powers of $1/\log s$. Asymptotically in $s\to \infty$, only the zeroth moment survives and the form factor is fully determined in terms of $f_\pi$ and $\alpha_s(s)$, see e.g.\ \cite{Sterman:1997sx}. This is the high-energy behavior used in the recent SVZ-inspired bootstrap \cite{He:2023lyy,He:2024nwd,He:2025gws}. However, it is known that the logarithmic decay of the moments might be \textit{very slow}, making said asymptotic form unreliable at the scales that they work with. In this paper we will only exploit the overall suppression power $F(s) = O(s^{-1})$, and remain agnostic about its coefficient. We finish by mentioning that the modern description of the factorization \eqref{eq:Ffactorization} as well as perturbative and non-perturbative corrections thereof uses soft-collinear effective theory (SCET) \cite{Bauer:2002nz}, but a discussion on that topic would take us too far afield.

\subsubsection{Scattering amplitude}
For the scattering amplitude $M(s,u)$, there are different high-energy limits depending on how we scale $u$. As usual, we will consider the standard Regge limit; $|s|\to \infty$ with $u\lesssim 0$ fixed. This limit is controlled by Regge theory (see e.g.\ appendix A in \cite{Albert:2023jtd} for a review), which predicts that the amplitude behaves as
\begin{equation}\label{eq:Regge}
  \lim_{|s|\to \infty} M(s,u)\sim  s^{\alpha_0(u)} \,,
\end{equation}
with $\alpha_0(u)$ the \textit{leading Regge trajectory}. For Large $N$ QCD, this corresponds to the trajectory of the rho meson, which intercepts the $u=0$ axis below 1, implying that
\begin{equation}\label{eq:Regge2}
    \lim_{|s|\to \infty} \frac{M(s,u)}{s}=0 \,.
\end{equation}
A similar statement holds for $M(s,-s-u)$.
Let us point out that, unlike the form factor and two-point function, this high-energy limit is not directly controlled by the UV theory (and its asymptotic freedom) because there remains a finite scale $u$. Instead, the limit that is controlled by UV QCD is the fixed-angle high-energy limit, $|s|\to \infty$ with $u/s$ fixed, leading to the celebrated Brodsky-Farrar counting rules \cite{Brodsky:1973kr,Lepage:1980fj}.

\section{Rigorous EFT bounds}\label{sec:rigorousBounds}
With the setup and definitions out of the way, we can proceed with the bootstrap. In this section, we follow the standard logic of positivity bounds to place \textit{rigorous bounds} on low-energy coefficients. We leave for section \ref{sec:PhenoBounds} stronger bounds that follow from asymptotic freedom, which, as we will explain, are necessarily phenomenological in nature.

\subsection{Sum rules from dispersion relations}
As discussed in section \ref{sec:HEbehavior}, the high energy behaviors of $\Pi(s)$, $F(s)$, and $M(s,u)$ are all polynomially bounded.
This allows us to write down dispersion relations that kill the contribution of a contour running around infinity,
\begin{alignat}{2}\label{eq:InftyLoops}
    &\frac{1}{2\pi i}\oint_\infty ds\, \frac{\Pi(s)}{s^{k+1}} = 0\,, \qquad
    &&\frac{1}{2\pi i}\oint_\infty ds\, \frac{F(s)}{s^{k}} = 0\,,\\
    &\frac{1}{2\pi i } \oint_\infty ds\, \frac{M(s, u)}{s^{k+1}}=0\,, \qquad 
    &&\frac{1}{2\pi i } \oint_\infty ds\, \frac{M(s, -s-u)}{s^{k+1}}=0\,,\nonumber
\end{alignat}
where $k=1,2,...$. The bottom two relations are respectively the SU and ST dispersion relations discussed in \cite{Albert:2022oes}; the ``new'' ingredients for the bootstrap are the relations in the top line.\footnote{Dispersion relations for form factors, and especially for two-point functions, are of course not new. They date back to the K\"all\'en-Lehmann representation \cite{Kallen:1952zz,Lehmann:1954xi} and Weinberg sum rules \cite{Weinberg:1967kj}.} By shrinking the contours towards the real axis and plugging the low-energy expansions \eqref{eq:lowEexp} around $s\sim 0$, we can derive sum rules for the low-energy coefficients of the propagator and the form factor,
\begin{gather}\label{eq:ab-sumrules}
    a_k = \int_{M^2}^\infty dm^2\, \rho^{(\gamma\gamma)}(m^2) \frac{1}{m^{2k+2}}\,, \qquad
    b_{k-1} = \int_{M^2}^\infty dm^2\, \rho^{(\gamma2\pi)}(m^2) \frac{1}{m^{2k}}\,.
\end{gather}
For the amplitude, we need to further take derivatives in $u$, and we find the sum rules \cite{Albert:2022oes}
\begin{align}
    g_{n,\ell} =&\, \sum_J \frac{n_J}{\pi}\int_{M^2}^\infty \frac{dm^2}{m^2} \rho_J^{(4\pi)}(m^2)\frac{2^{\ell-\delta_{n,2\ell}}}{\ell !}\frac{\mathcal P_J^{(\ell)}(1)}{m^{2n}}\, ,\qquad
    \begin{aligned}
        n &= 1, 2, \dots \, \\
        \ell&=0,\dots,\left[\tfrac{n}{2}\right]\,.
    \end{aligned}
\end{align}
On top of these sum rules, one finds two infinite towers of \textit{null constraints} encoding the crossing symmetry of the amplitude \cite{Tolley:2020gtv,Caron-Huot:2020cmc}. We refer the reader to \cite{Albert:2022oes} for their precise expressions. No null constraints come from $\Pi(s)$ and $F(s)$ because they have no non-trivial crossing properties.

Following \cite{Caron-Huot:2020cmc}, it is convenient to introduce high-energy averages to simplify notation. We define
\begin{equation}
    \Big<\left(\cdots\right)\Big>_{J\neq 1} \equiv \frac{1}{\pi} \sum_{J\neq 1} n_J\int_{M^2}^\infty\frac{dm^2}{m^2} \rho_J^{(4\pi)}(m^2) \left(\cdots\right)\,,
\end{equation}
which averages over all the contributions to the $\pi\pi \to \pi \pi$ amplitude of spin different from one. For the spin-one states, we define an average combining the contributions to the amplitude with those to the propagator and the form factor,
\begin{equation}\label{eq:J=1avg}
    \left<\begin{pmatrix}
        \cdots &  & \cdots\\
        \cdots & & \cdots
    \end{pmatrix}\right>_{J= 1} \equiv \int_{M^2}^\infty\frac{dm^2}{m^2} \text{Tr}\left(\begin{pmatrix}
        \rho^{(\gamma\gamma)}(m^2) & \frac{1}{\sqrt{2}}\rho^{(\gamma 2\pi)}(m^2)\\
        \frac{1}{\sqrt{2}}\rho^{(\gamma 2\pi)}(m^2) & \frac{n_1}{\pi}\rho_1^{(4\pi)}(m^2)
    \end{pmatrix} \begin{pmatrix}
        \cdots &  & \cdots\\
        \cdots & & \cdots
    \end{pmatrix}\right)\,.
\end{equation}
The crucial feature of these averages is that they have a \textit{positive measure} as a consequence of unitarity. Indeed $\rho_J^{(4\pi)}(m^2)\geq 0$, and the kernel in \eqref{eq:J=1avg} is $\tilde \rho_1(m^2)$, which we showed is a positive semidefinite matrix in \eqref{eq:rho1}. This implies that $\langle\cdots \rangle_{J\neq 1}$ returns a positive number if its argument is positive, and $\langle\cdots \rangle_{J = 1}$ does so if its argument is a positive-semidefinite matrix.

To assist the reader, we write explicitly some of the sum rules (in bracket notation) that will be relevant below:
\begin{alignat}{2}\label{eq:SRs}
    a_1 =&\, \left< \begin{pmatrix}
        \frac{1}{m^2} & 0 \\
        0 & 0
    \end{pmatrix}\right>_{J=1}\,,
    \qquad
    && \hspace{6pt}a_2 = \left< \begin{pmatrix}
        \frac{1}{m^4} & 0 \\
        0 & 0
    \end{pmatrix}\right>_{J=1}\,, \\
    b_0 =&\, \left< \begin{pmatrix}
        0 & \frac{1}{\sqrt{2}} \\
        \frac{1}{\sqrt{2}} & 0
    \end{pmatrix}\right>_{J=1}\,, \qquad
    &&g_{1,0} = \left< \begin{pmatrix}
        0 & 0 \\
        0 & \frac{1}{m^2}
    \end{pmatrix}\right>_{J=1} + \left< \frac{1}{m^2}\right>_{J\neq 1}\,, \nonumber\\
    b_1 =&\, \left< \begin{pmatrix}
        0 & \frac{1}{\sqrt{2}\,m^2} \\
        \frac{1}{\sqrt{2}\,m^2} & 0
    \end{pmatrix}\right>_{J=1}\,, \qquad
    &&g_{2,0} = \left< \begin{pmatrix}
        0 & 0 \\
        0 & \frac{1}{m^4}
    \end{pmatrix}\right>_{J=1} + \left< \frac{1}{m^4}\right>_{J\neq 1}\,,\nonumber\\
    b_2 =&\, \left< \begin{pmatrix}
        0 & \frac{1}{\sqrt{2}\,m^4} \\
        \frac{1}{\sqrt{2}\,m^4} & 0
    \end{pmatrix}\right>_{J=1}\,,
    \qquad
    &&g_{2,1} = \left< \begin{pmatrix}
        0 & 0 \\
        0 & \frac{1}{m^4}
    \end{pmatrix}\right>_{J=1} + \left< \frac{J(J+1)}{2m^4}\right>_{J\neq 1}\,.\nonumber
\end{alignat}

Since null constraints are only present in the four-point amplitude, they all take the form
\begin{equation}\label{eq:NCs}
    \left< \begin{pmatrix}
        0 & 0 \\
        0 & \mathcal N_{i}(m^2,1)
    \end{pmatrix}\right>_{J=1} + \left< \mathcal N_{i}(m^2,J)\right>_{J\neq 1} = 0\,,
\end{equation}
with $\mathcal N_i(m^2,J)$ a placeholder for the null constraints derived in \cite{Albert:2022oes}.

\subsection{Exclusion plot}\label{sec:exclusionPlot}
The simplest bounds one can derive in this system concern the propagator alone. Clearly, \eqref{eq:ab-sumrules} implies that all $a_k$ are positive, but one can also recognize an upper bound on the ratio of couplings by noting that the integral in \eqref{eq:ab-sumrules} only has support above the cutoff $M^2$. It follows that
\begin{equation}
    0 \leq \frac{a_k M^{2(k-j)}}{a_j} \leq 1\,, \qquad 1\leq j\leq k\,.
\end{equation}
These bounds constrain the counterterms in the chiral Lagrangian generated by the RG flow (recall the discussion at the end of section \ref{sec:LEbehavior}). It would be interesting to test them against real-world data, but we are not aware of experimental/lattice determinations of these coefficients.

The form factor couplings $b_i$, on the other hand, only populate off-diagonal entries of the sum rules \eqref{eq:SRs}, so it is not possible to bound these coefficients without invoking the full mixed system. The trick is to look for linear combinations whose sum rules combine into the average of a positive semidefinite matrix. For example, with $a_1, b_1, g_{1,0}$ we can construct
\begin{equation}\label{eq:xyz-arg}
    x a_1 + y b_1 + z g_{1,0} = \left< \begin{pmatrix}
        \frac{x}{m^2} & \frac{y}{\sqrt{2}\,m^2} \\
        \frac{y}{\sqrt{2}\,m^2} & \frac{z}{m^2}
    \end{pmatrix}\right>_{J=1} + \left< \frac{z}{m^2}\right>_{J\neq 1}\,.
\end{equation}
The right hand side is positive as long as $x,z\geq 0$ and $x z - y^2/2 \geq 0$. The optimal (nontrivial) bound comes from the saturation of the last inequality. By choosing $x=1/a_1$ and $z=1/g_{1,0}$, we derive the bound
\begin{equation}\label{eq:boundonb1}
    |\tilde b_1| \leq \sqrt{2}\,, \qquad \text{where}\quad \tilde b_1 \equiv \frac{b_1}{\sqrt{a_1 g_{1,0}}}\,.
\end{equation}

A similar strategy can be followed to place bounds on the other normalized couplings
\begin{equation}
    \tilde{b}_i \equiv \frac{b_i M^{2(i-1)}}{\sqrt{a_1 g_{1,0}}} \,,
\end{equation}
except for $\tilde b_0$, which comes at lower mass order than $a_1,g_{1,0}$. The optimal bounds for $i>1$ make use of the null constraints \eqref{eq:NCs}, so it becomes useful to streamline the search for bounds using a semidefinite program solver such as \texttt{SDPB} \cite{Simmons-Duffin:2015qma}. The algorithm is by now standard (see e.g.\ \cite{Caron-Huot:2020cmc,Albert:2022oes,Albert:2023jtd}), so we will refrain from repeating it here. Including null constraints up to Mandelstam order $n_\text{max}=8$,\footnote{This corresponds to 28 null constraints. See \cite{Albert:2022oes} for the precise definition of $n_\text{max}$.
} we find
\begin{equation}
    |\tilde{b}_2|\leq 1.246334... \,,
\end{equation}
which converges quite fast with the number of null constraints. All the bounds on $|\tilde b_k|$ with $k\geq 2$ turn out to be numerically the same.

\begin{figure}
    \centering
    \includegraphics[width=0.6\linewidth]{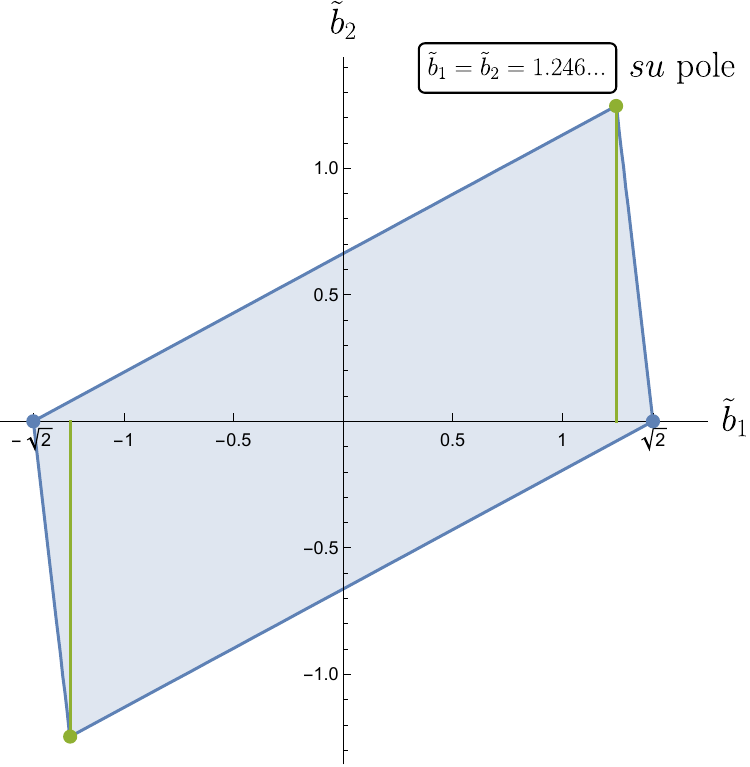}
 \caption{Exclusion plot in the space of normalized form factor couplings $\tilde{b}_1$ and $\tilde{b}_2$ with $n_{\text{max}}=5$. The green vertical lines are ruled in by the $su$-\textit{pole} amplitude.}
    \label{fig:exclusionplot}
\end{figure}

Running the program to get a bound on $\tilde{b}_2$ for various fixed values of $\tilde{b}_1$, we generate the exclusion plot shown in figure \ref{fig:exclusionplot}. Since the sign of the form factor spectral density is unconstrained, the allowed region is symmetric with respect to a simultaneous sign flip in both axes. We see that it forms a convex polygon, as it should. Therefore, identifying the theories that live on the corners of the plot is sufficient to analytically rule in the entire allowed region and understand the nature of the bounds. The left and right corners of the plot correspond to a theory in which all masses in the spectrum are pushed to infinity. This follows from dimensional analysis; $\tilde b_2$ is proportional to $M^2$ but $\tilde b_1$ is not, so in the limit of $M\to \infty$, the bound on the former drops to zero, while the point $\tilde b_1 = \sqrt{2}$ remains. It is not immediately obvious precisely which family of amplitudes saturates this bound, but we deem it uninteresting due to its lack of finite-energy states.

The remaining corners are kinks at $\tilde{b}_1=\tilde{b}_2\simeq \pm1.246$. Drawing inspiration from \cite{Caron-Huot:2020cmc,Albert:2022oes}, a natural candidate for the ``theory'' residing at these kinks is an infinite tower of states with increasing spin $J\geq 1$ and equal mass. The corresponding amplitude takes the form
\begin{equation}\label{eq:inftowerampl}
    M_{su-\text{pole}}(s,u) = \frac{m^4}{(m^2-s)(m^2-u)} - \alpha_{0}\left(\frac{m^2}{m^2-s}+ \frac{m^2}{m^2-u} \right)  \;.
\end{equation}
The residue at $s=m^2$ of the first term can be projected onto the orthogonal set of Legendre polynomials to separate the contributions coming from each spin state. Since it is not polynomial in $u$, all spins receive contributions. The second term in \eqref{eq:inftowerampl} is a spin-zero amplitude with $\alpha_0$ tuned to cancel out the spin-zero contribution of the first term; $\alpha_0 = \log 2$.

Since this amplitude has a single spin-one state, from \eqref{eq:spin1exchange} we know that the extension to the two-point function and form factor is simply
\begin{align}
    \Pi(s) = \frac{f_\rho^2}{m^2 - s}\,,\qquad 
    F(s) =  \frac{g_{\pi\pi \rho}f_\rho m}{m^2 - s}\,,
\end{align}
and the corresponding low-energy coefficients,
\begin{equation}
    a_1 = \frac{f_\rho^2}{m^4}, \quad b_1 = \frac{g_{\pi\pi\rho} f_\rho}{m^3}, \quad b_2 = \frac{g_{\pi\pi\rho} f_\rho}{m^5}, \quad g_{1,0} = \frac{1-\alpha_0}{m^2} \;.
\end{equation}
The pion$-$rho coupling $g_{\pi\pi\rho}$ is fixed by extracting the spin-one component \eqref{eq:spin1amplitude} from \eqref{eq:inftowerampl},
\begin{equation}
    g_{\pi\pi\rho}^2 = 2\log 512 - 12\,.
\end{equation}
The current$-$rho coupling $f_\rho$ can remain unfixed. The normalized couplings are then
\begin{equation}
     \tilde{b}_1 = \pm\sqrt{\frac{2\log 512 - 12}{1-\log 2}}\simeq \pm 1.2463... \, , \quad  \tilde{b}_2 = \pm\sqrt{\frac{2\log 512 - 12}{1-\log 2}}\left(\frac{M^2} {m^2} \right)\,.
\end{equation}
This defines two vertical lines (drawn in green in figure \ref{fig:exclusionplot}) as a function of $m/M$, which rule in the kinks. Positive linear combinations of these solutions and the infinite-mass theory then analytically rule in the entire allowed region. Neither of these corner theories are physically meaningful, but they are compatible with the bootstrap assumptions.

\subsection{Including the rho}
\begin{figure}
    \centering
    \includegraphics[width=0.95\linewidth]{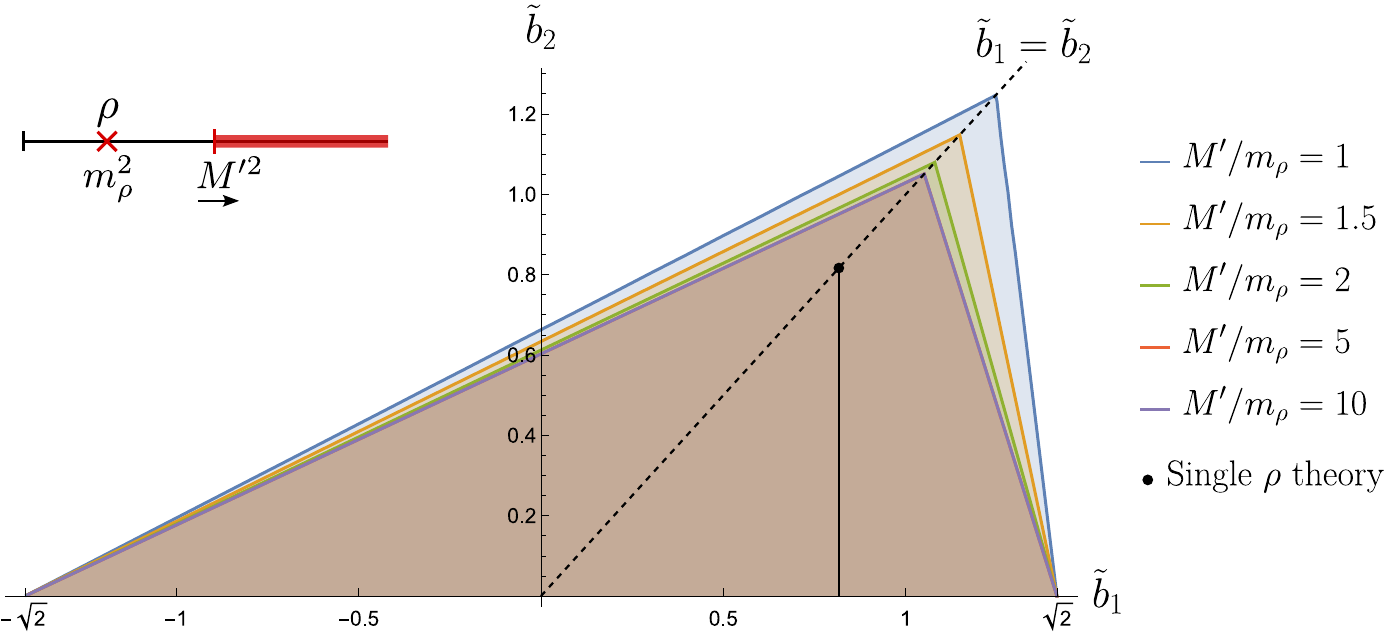}
 \caption{Exclusion plot in the upper half plane of normalized form factor couplings $\tilde{b}_1$ and $\tilde{b}_2$ with the rho meson included in the spectrum, for several values of the cutoff $M'/m_\rho$, with $n_{\text{max}}=7$. The black line is ruled in by the single rho exchange, descending vertically from $\tilde{b}_1=\tilde{b}_2=\sqrt{2/3}$. The numerical kinks are expected to converge to this theory as more null constraints are included.}
    \label{fig:exclusionplot-rho}
\end{figure}

We can strengthen the bounds by imposing spectral assumptions that bring us closer to large $N$ QCD. The first step in this direction is to integrate back into the EFT the lowest-lying meson, the rho. We take a spin-one particle at mass $m_\rho=M$, and all other states starting no earlier than a new cutoff $M'$. In principle, one should redo the dispersion relation analysis using the refined EFT at energies below $M'$ to derive new sum rules and null constraints. As pointed out in \cite{Albert:2022oes}, however, one can formally keep the rho as a high-energy state and simply assume a gap between it and the higher states. The corresponding exclusion plot for various values of the ratio $M'/m_\rho$ is given in figure \ref{fig:exclusionplot-rho}. We see that the $su$-pole kink gets excluded, as expected, since that amplitude has an infinite tower of spins degenerate with the rho. The kink swiftly moves down and converges to another point as the new cutoff is pushed to infinity.

In this limit, we probe theories that only contain the rho meson exchange. A single spin-one exchange is not compatible with the better-than-one Regge behavior \eqref{eq:Regge2}, but the fact that a nontrivial kink survives this limit points to the existence of a family of allowed solutions which tend to the single rho amplitude in the limit of $M'/m_\rho\to \infty$. A pion amplitude for such a family was found in \cite{Albert:2022oes},
\begin{equation}
M_{\rho}(s,u)=\frac{m_\rho^2+2 u}{m_\rho^2-s}\left(\frac{m_{\infty}^2}{m_{\infty}^2-u}\right)+\frac{m_\rho^2+2 s}{m_\rho^2-u}\left(\frac{m_{\infty}^2}{m_{\infty}^2-s}\right) \,,
\end{equation}
where $m_\infty$ is an arbitrarily large (but finite) auxiliary mass. This amplitude has spin-one states at $m_\rho$ and at $m_\infty$, so the corresponding two-point function and form factor are
\begin{equation}
    \Pi(s) = \frac{f_\rho^2}{m_\rho^2 - s} + \frac{f_\infty^2}{m_\infty^2 - s}\,, \qquad 
    F(s) = \frac{g_{\pi\pi\rho} f_\rho m_\rho}{m_\rho^2 - s} + \frac{g_{\pi\pi\infty} f_\infty m_\infty}{m_\infty^2 - s}\,.
\end{equation}
The three-point couplings are extracted from the amplitude,
\begin{align}
    g_{\pi\pi\rho}^2 =&\, 6\left(\left(1+\frac{2m_\infty^2}{m_\rho^2}\right)\log\left(1+\frac{m_\rho^2}{m_\infty^2}\right)-2\right)\frac{m_\infty^2}{m_\rho^2}\left(1+\frac{2m_\infty^2}{m_\rho^2}\right)\,,\\
    g_{\pi\pi\infty}^2 =&\, 6\left(\left(1+\frac{2m_\rho^2}{m_\infty^2}\right)\log\left(1+\frac{m_\infty^2}{m_\rho^2}\right)-2\right)\left(2+\frac{m_\rho^2}{m_\infty^2}\right)\,.\nonumber
\end{align}
We note that, despite appearances, the first of these couplings tends to a constant in the limit of $m_\infty/m_\rho \to \infty$, approaching a theory with a single spin-one meson.

These functions satisfy the high energy behavior assumed in \eqref{eq:InftyLoops} and the unitarity condition \eqref{eq:rho1} for any value of the couplings $f_\rho,f_\infty$, so they should fall within the allowed region of the exclusion plots. Computing their low-energy expansions and taking the limit of $m_\infty \to \infty$ with finite $f_\rho,f_\infty$, we find 
\begin{equation}
    \tilde b_1 = \pm \sqrt{\frac{2}{3}} \simeq 0.8165...\,, \qquad  \tilde b_2 = \pm \sqrt{\frac{2}{3}} \left(\frac{M^2}{m_\rho^2}\right)\,.
\end{equation}
This rules in the vertical line drawn in black in figure \ref{fig:exclusionplot-rho}. While it is currently well within the bounds, we expect it to describe the kink in the limit of $n_\text{max}\to\infty$. With every increase of $n_\text{max}$, the value of $\tilde b_1,\tilde b_2$ to which the kink converges for $M'/m_\rho\to \infty$ gets lower, but the convergence is rather slow. The same phenomenon was observed for certain kink in \cite{Albert:2022oes}, and it was later shown analytically that the numerical kink would go down to meet the spin-one amplitude \cite{Fernandez:2022kzi}. It would be interesting to repeat such an argument in the current case.

\subsection{Including the \texorpdfstring{\textit{f}\textsubscript{\hspace{-1pt}2}}{f2}}\label{sec:including-f2}
The natural refinement to the spectral assumption of the previous section is to include one more meson after the rho. At large $N$, the lightest candidate for this is the $f_2$ meson, and the pion amplitude under this spectral assumption was investigated in \cite{Albert:2023seb}.  It was found that the upper bound on the coupling $g_{\pi\pi f_2}$, as a function of the spectral gap $\tilde{M}$ after the $f_2$ pole, exhibits a sharp kink at $\tilde{M}\sim 2.18\; m_\rho$, which approximately corresponds to the real-world mass of the $\rho_3$ meson. The extremal solution at this $f_2$ \textit{kink} contains a full Regge trajectory; we refer the reader to \cite{Albert:2023seb} for a detailed discussion. In this section, we investigate the imprints of the $f_2$ kink on the form factor couplings.

\begin{figure}
    \centering
    \includegraphics[width=0.93\linewidth]{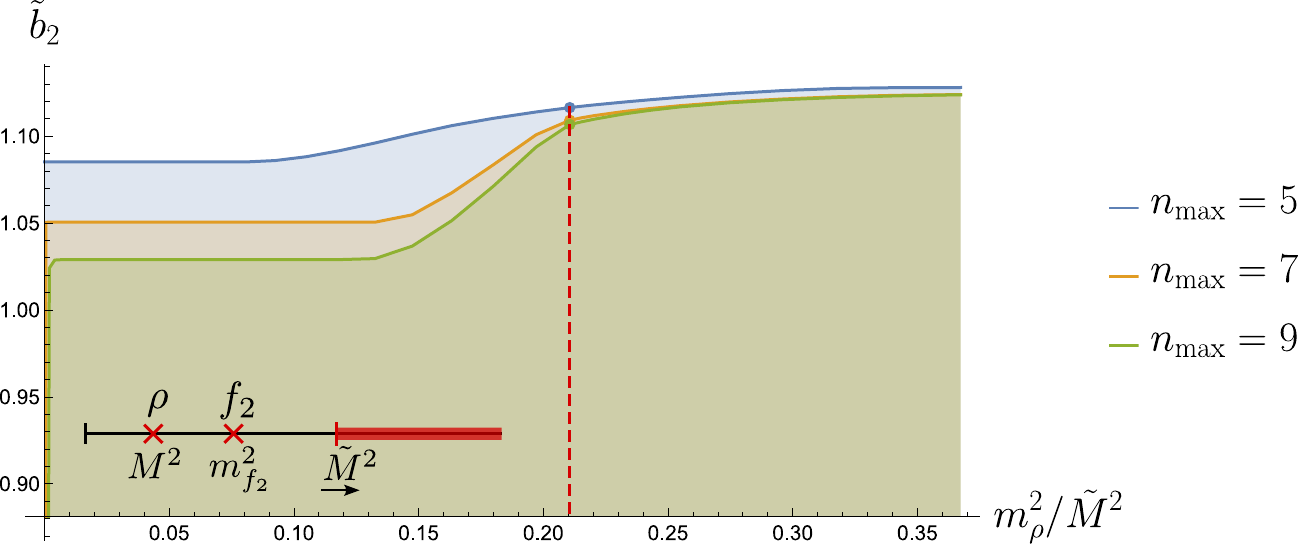}
 \caption{The upper bound on $\tilde{b}_2$ as a function of the ratio $m_\rho^2/\tilde{M}^2$ with the outlined spectral input, for different numbers of null constraints $n_{\text{max}}$. The vertical red line marks the location of the kink identified in \cite{Albert:2023seb}. }
    \label{fig:b2vsMtilde-withf2}
\end{figure}

The ratio of masses of the first two mesons in the spectrum is fixed to a value borrowed from real-world QCD,
\begin{equation}
m_{f_2} \simeq 1.65\, m_\rho \,,
\end{equation}
and a new cutoff $\tilde{M}$ is assumed above the $f_2$ pole as a free parameter. The bound on $\tilde b_1$ with these spectral assumptions is again \eqref{eq:boundonb1} and
has no dependence on $m_\rho^2/\tilde M^2$. This is because the bound is saturated by a solution where the rho and $f_2$ couplings are turned off and all other states are pushed to infinity. Instead, the bound on $\tilde b_2$ has a nontrivial dependence on the ratio $m_\rho^2/\tilde{M}^2$, shown in figure \ref{fig:b2vsMtilde-withf2}.

As we increase the cutoff smoothly from $m_{f_2}$, the bound initially changes slowly until it reaches a critical value where it quickly drops to an asymptotic plateau reaching $\tilde M\to \infty$. This sudden change of behavior occurs unsurprisingly around the location of the $f_2$ kink~\cite{Albert:2023seb}, marked with a red dashed line in figure \ref{fig:b2vsMtilde-withf2}. Higher values of $n_\text{max}$ would undoubtedly delineate a sharper kink at that point. Taking larger values of $\tilde{M}$ produces the plateau on the left of the plot, whose value corresponds to the numerical bound that the bootstrap finds for $\tilde{b}_2$ in a theory containing a single rho. This is expected: as the new cutoff $\tilde{M}$ is pushed higher, the bootstrap can only find a valid solution by setting the coupling of the $f_2$ meson to zero, leaving the extremal solution with a single rho.

\section{Phenomenological bounds}\label{sec:PhenoBounds}
We have seen that positivity bounds on large $N$ pion scattering amplitudes can be extended into a system involving the form factor and vector current two-point function. While this system granted us access to new interesting observables of the low-energy theory, it did not lead to any new solutions to the bootstrap beyond those found directly with the scattering amplitude in \cite{Albert:2022oes,Albert:2023jtd,Albert:2023seb,Fernandez:2022kzi,Ma:2023vgc,Li:2023qzs,Eckner:2024pqt,Berman:2024kdh}. The reason is that $\Pi(s)$ and $F(s)$ did not introduce any new relations constraining on-shell data further. The true worth of this system lies in the fact that the high-energy behavior of $\Pi(s)$ is under perturbative control thanks to asymptotic freedom of the underlying gauge theory, as we reviewed in section \ref{sec:HEbehavior}. This allows for the implementation of new dispersion relations where the contribution from the arc at infinity is computed with the UV behavior rather than killed with a suitable kernel. Such relations, which we dub SVZ-like sum rules after \cite{Shifman:1978bx,Shifman:1978by}, let us feed explicit information from QCD into the form factor bootstrap. This strategy was pointed out and executed in \cite{He:2023lyy,He:2024nwd,He:2025gws}, in the context of the (finite~$N$) non-perturbative $S$-matrix bootstrap \cite{Paulos:2016fap,Paulos:2016but,Paulos:2017fhb,Homrich:2019cbt,Guerrieri:2018uew,Guerrieri:2020bto}. Here we discuss the setup and consequences in the large $N$ setting.

\subsection{SVZ-like sum rules}
Let us recall the asymptotic result \eqref{eq:PipQCD} from leading order perturbation theory:
\begin{equation}\label{eq:1loop}
    \Pi_{\text{pQCD}}(s) \simeq -\frac{N}{24\pi^2} \log(-s/\mu^2) + \cdots\,.
\end{equation}
As discussed, instead of taking a kernel $K(s)\sim 1/s$ to kill the high-energy arc in dispersion relations, we now want to keep it explicitly. Ideally, we would like to use a kernel, like $K(s) \sim 1/\log(-s/\mu^2)$, which singles out the coefficient of the $\log$ in $\oint_{\infty} ds\, K(s) \Pi(s) /s$. Unfortunately, any such kernel appears to introduce a whole ``subtraction cut'', rather than a point, along which we would need to specify $\Pi(s)$. This hinders our prospects of inputting the high-energy behavior into the bootstrap. One way to make progress is to give up the mathematical rigor of the bounds from the previous section in favor of approximate bounds of somewhat more phenomenological flavor.

\begin{figure}[t]
    \centering
    \includegraphics[width=\textwidth]{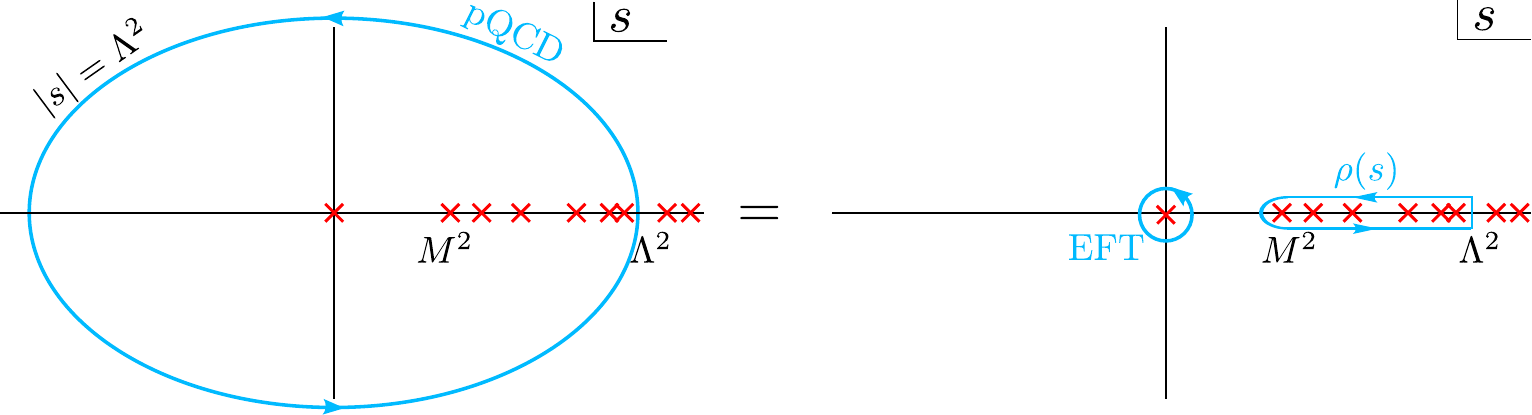}
    \caption{Contour deformation used to derive the SVZ-like sum rules. $M^2$ denotes the cutoff of the low-energy EFT, and $\Lambda^2$ the scale at which asymptotic freedom kicks in and the perturbative result can be used. 
    \label{fig:svzanalyticplane}
    }
\end{figure}

The idea, dating back to the late 70s with work of Shifman, Vainshtein and Zakharov (SVZ) \cite{Shifman:1978bx,Shifman:1978by}, is to assume that there is a scale $|s|\sim \Lambda^2$ at which the one-loop result \eqref{eq:1loop} is already a good approximation of the full $\Pi(s)$, so that we can use $\Pi_{\text{pQCD}}(s)$ to evaluate contour integrals around that scale. By shrinking the contour to pick up the low-energy contribution and imaginary parts, as in figure \ref{fig:svzanalyticplane}, we obtain the UV-IR relation
\begin{equation}\label{eq:IR-UVsumrules}
    \frac{1}{2\pi i } \oint_{0} ds \; \frac{\Pi_{\text{low}}(s)}{s^{k+1}}-\frac{1}{2\pi i }\oint_{\Lambda^2} ds \; \frac{\Pi_{\text{pQCD}}(s)}{s^{k+1}} = \frac{1}{\pi}\int_{M^2}^{\Lambda^2} ds \; \frac{\text{Im}\,\Pi(s)}{s^{k+1}} \;.
\end{equation}
For $k\geq 1$, we can take the limit $\Lambda\to \infty$ and recover the standard sum rules from section~\ref{sec:rigorousBounds}. If we keep $\Lambda$ finite, the UV contribution converges, and we can take any value of $k$. This allows us to derive sum rules for coefficients that were inaccessible before. For instance, for $k=0$ we have
\begin{equation}\label{eq:a0SVZ}
    a_0 +\frac{N}{24\pi^2}\log(\Lambda^2/\mu^2) = \int_{M^2}^{\Lambda^2}\frac{dm^2}{m^2}\, \rho^{(\gamma\gamma)}(m^2)\,.
\end{equation}
While morally equivalent, the traditional SVZ sum rules differ from these in their technical implementation. We include a short review of the latter in appendix \ref{appendix:svz}, for the benefit of the reader unfamiliar with the subject.

The intuition behind the sum rule \eqref{eq:a0SVZ} is clear. The current-current spectral density asymptotes to a constant, $\rho^{(\gamma\gamma)}(m^2)\to N/24\pi^2$, as discussed around \eqref{eq:asymprho}. Thus, in the limit of $\Lambda^2\to \infty$, the right-hand side of \eqref{eq:a0SVZ} diverges, preventing us from capturing the $a_0$ constant. The assumption that the asymptotic behavior is reached at a finite energy $\Lambda$ allows us to repackage the UV contribution and truncate the diverging integral, producing a sum rule. We also note that \eqref{eq:a0SVZ} makes explicit the cancellation of counterterms discussed at the end of section \ref{sec:LEbehavior}. We can always tune the scheme dependence of $\Pi_{\text{pQCD}}(s)$ by shifting an overall constant in $\Pi(s)$, but this cancels out between the UV and IR contributions, leaving the sum rule unchanged.

The reason why we insist on calling these sum rules \textit{phenomenological} is that they are only approximately correct. They can be systematically improved by including subleading corrections to the high-energy behavior \eqref{eq:1loop}. It follows from $\eqref{eq:PipQCD}$ that perturbative corrections to $\Pi_{\text{pQCD}}(s)$ are of order $O(\log(-s)/\log\Lambda^2)$, while non-perturbative corrections kick in at order $O(1/s^2)$.
This translates into errors of the sum rule \eqref{eq:a0SVZ} of order $O(\Lambda^0)$.
For higher-subtracted dispersion relations, $k<0$, the sum rules (and their errors) read
\begin{equation}
    a_k - \frac{N}{24\pi^2}\frac{1}{ k\Lambda^{2k}} = \int_{M^2}^{\Lambda^2}\frac{dm^2}{m^2}\, \rho^{(\gamma\gamma)}(m^2)\frac{1}{m^{2k}} + O\left(\frac{\Lambda^{-2k}}{\log \Lambda^2}\right) \,, \qquad k=1,2,...\,.
\end{equation}
For anti-subtracted dispersion relations, $k=-\ell<0$, instead
\begin{equation}\label{eq:a-1UV}
    \frac{N}{24\pi^2}\frac{\Lambda^{2\ell}}{\ell} = \int_{M^2}^{\Lambda^2}\frac{dm^2}{m^2}\, \rho^{(\gamma\gamma)}(m^2)m^{2\ell} + O\left(\frac{\Lambda^{2\ell}}{\log \Lambda^2}\right)\,, \qquad \ell=1,2,...\,.
\end{equation}
Notice that the situation here is completely analogous to the case of positivity bounds with EFT loops (see e.g.\ \cite{Bellazzini:2020cot,Bellazzini:2021oaj,Beadle:2024hqg,Chang:2025cxc,Beadle:2025cdx}), where to compute ``IR arcs'' one needs to assume the existence of a low-energy scale at which the $n$-loop amplitude is a good approximation of the full result.

Before jumping to the bounds that we can achieve with these new sum rules, we need to apply a similar SVZ-like treatment to the form factor $F(s)$. We assume that the high-energy result $F(s)\sim s^{-1}$, discussed in section \ref{sec:FormFactorHE}, is already a good approximation at energies around $s\sim \Lambda^2$, and we consider a contour integral around that scale. Recall that the coefficient of this leading power is a series of log-suppressed moments of the pion distribution amplitudes.\footnote{In \cite{He:2023lyy,He:2024nwd,He:2025gws}, they assume a high-energy behavior for $F(s)$ given solely by the zeroth moment, which is known analytically.
While it is true that the coefficient of the $s^{-1}$ term eventually asymptotes to this result, the log decay can be very slow, so this behavior might settle in at a scale much larger than the perturbative QCD scale~$\Lambda$. We thank Simon Caron-Huot for discussions on this point.} Since these already include non-perturbative information, we will refrain from using antisubtracted dispersion relations. The UV contour then always vanishes at leading order. Performing the contour deformation from figure \ref{fig:svzanalyticplane} leads to the sum rules
\begin{equation}\label{eq:bkSVZ}
    b_k = \int_{M^2}^{\Lambda^2}\frac{dm^2}{m^2}\, \rho^{(\gamma2\pi)}(m^2)\frac{1}{m^{2k}} + O\left(\Lambda^{-2k-2}\right)\,, \qquad k=0,1,2,...\,.
\end{equation}
These are all sum rules that we already had access to before, but they are now truncated at a scale $\Lambda$ by the perturbative approximation (at the expense of an error).

\subsection{Bound on the pion decay constant}
We are now ready to derive new bounds from these sum rules. The first result that we will discuss is a bound on the pion decay constant, $f_\pi$, related to the coupling $g_{1,0}=1/(2f_\pi^2)$. This bound is a remarkable outcome of the form factor bootstrap, since this coupling is associated to the ``dimension six'' operator $(\pi\partial \pi)^2$, and traditional positivity methods only allow for bounds on ``dimension eight'' operators (normalized by the dimension six couplings).

\subsubsection{Bound on \texorpdfstring{\textit{b}\textsubscript{0}}{b0}}
The trick to achieve such a result is simply to place bounds on some ratio $b_k/\sqrt{a_\ell g_{1,0}}$ where both $b_k$ and $a_\ell$ are known. For the form factor, we use the first coupling of the low-energy expansion~\eqref{eq:lowEexp}, which is fixed to $b_0 = 1$ by symmetry requirements, recall \eqref{eq:Flow}.
For the two-point function, we consider the first antisubtracted dispersion relation,
\begin{equation}\label{eq:a-1SVZ}
    a_{-1} \equiv \frac{N}{24\pi^2}\Lambda^{2} = \avg{\begin{pmatrix}
        m^2 & 0\\ 0 & 0
    \end{pmatrix}}_{J=1}^{[M,\Lambda]} + O\left(\frac{\Lambda^2}{\log \Lambda^2}\right)\,,
\end{equation}
which is fully determined in terms of UV parameters of QCD. Note that we already had a sum rule for $b_0$ (given in \eqref{eq:SRs}), but it did not couple to any rigorous bounds because it had lower powers of $1/m^2$ than the normalization $\sqrt{a_1g_{1,0}}$.
Now that we have truncated sum rules, $b_0/\sqrt{a_{-1} g_{1,0}}$ will have a nontrivial bound.

\begin{figure}
    \centering
    \includegraphics[width=0.8\linewidth]{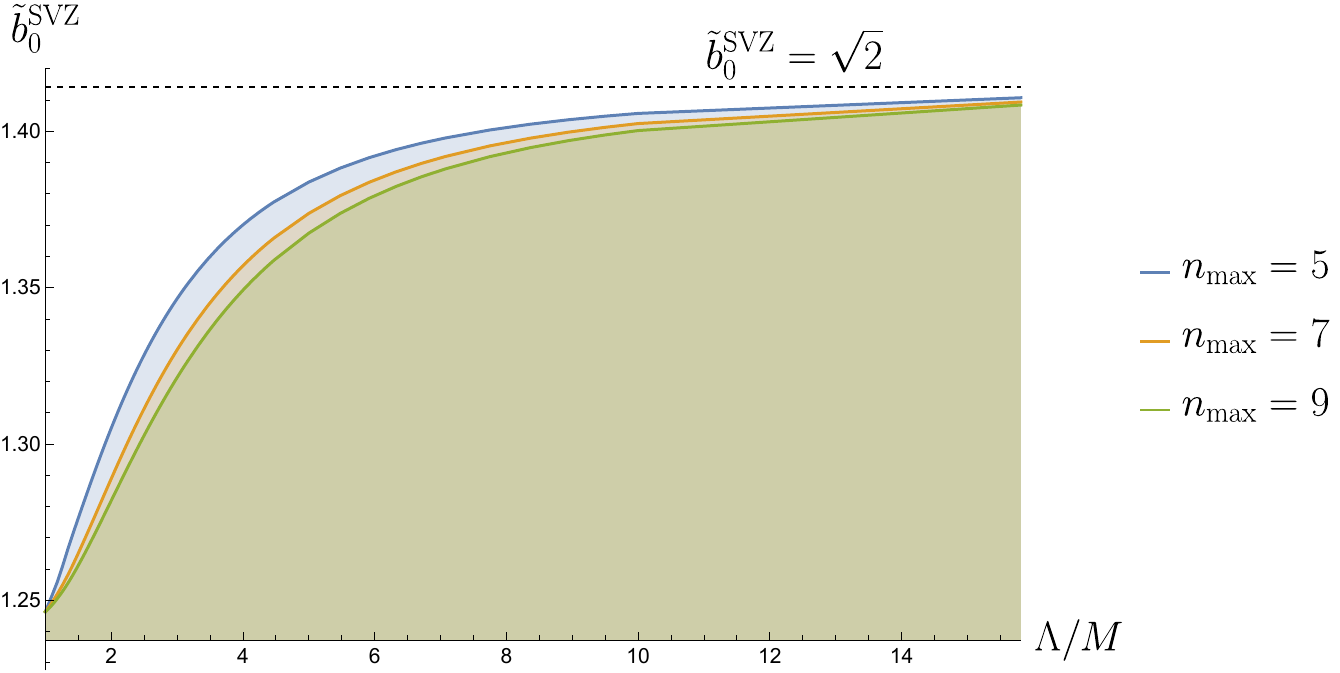}
    \caption{Bound on the normalized coupling $\tilde{b}_{0}^{\text{SVZ}}$ as $\Lambda/M$ is varied with various $n_{\text{max}}$ values.}
    \label{fig:boundonb0}
\end{figure}

Indeed, let us repeat the exercise from the beginning of section \ref{sec:exclusionPlot} for the current case. We consider the combination of sum rules
\begin{equation}\label{eq:xyz-arg2}
    x a_{-1} + y b_0 + z g_{1,0} = \left< \begin{pmatrix}
        \scriptstyle{x m^2} & \frac{y}{\sqrt{2}} \\
        \frac{y}{\sqrt{2}} & \frac{z}{m^2}
    \end{pmatrix}\right>_{J=1}^{[M,\Lambda]} + \left< \begin{pmatrix}
        0 & 0 \\
        0 & \frac{z}{m^2}
    \end{pmatrix}\right>_{J=1}^{(\Lambda,\infty)} + \left< \frac{z}{m^2}\right>_{J\neq 1}\,.
\end{equation}
Notice that, for the amplitude, we are including the contributions beyond $\Lambda$ because the Regge limit of $M(s,u)$ is not controlled by asymptotic freedom, and so the onset of its asymptotic behavior might be at a different scale. Positivity of the various arguments again leads to the simple bound
\begin{equation}\label{eq:b0SVZdef}
    \tilde b_0^{\text{SVZ}} \equiv \frac{b_0}{\sqrt{a_{-1} g_{1,0}}} \leq \sqrt{2}\,.
\end{equation}
It turns out that this bound is not optimal, though, and including null constraints into the mix strengthens it. Running a systematic semidefinite program with \texttt{SDPB} \cite{Simmons-Duffin:2015qma} for the upper bound on $\tilde b_0^{\text{SVZ}}$ as a function of $\Lambda/M$ produces figure \ref{fig:boundonb0}.\footnote{The only modification needed here with respect to the standard algorithm is to restrict the positivity requirement for the $J=1$ matrix to the domain $m\in [M,\Lambda]$.}

\begin{figure}
    \centering
    \includegraphics[width=0.8\linewidth]{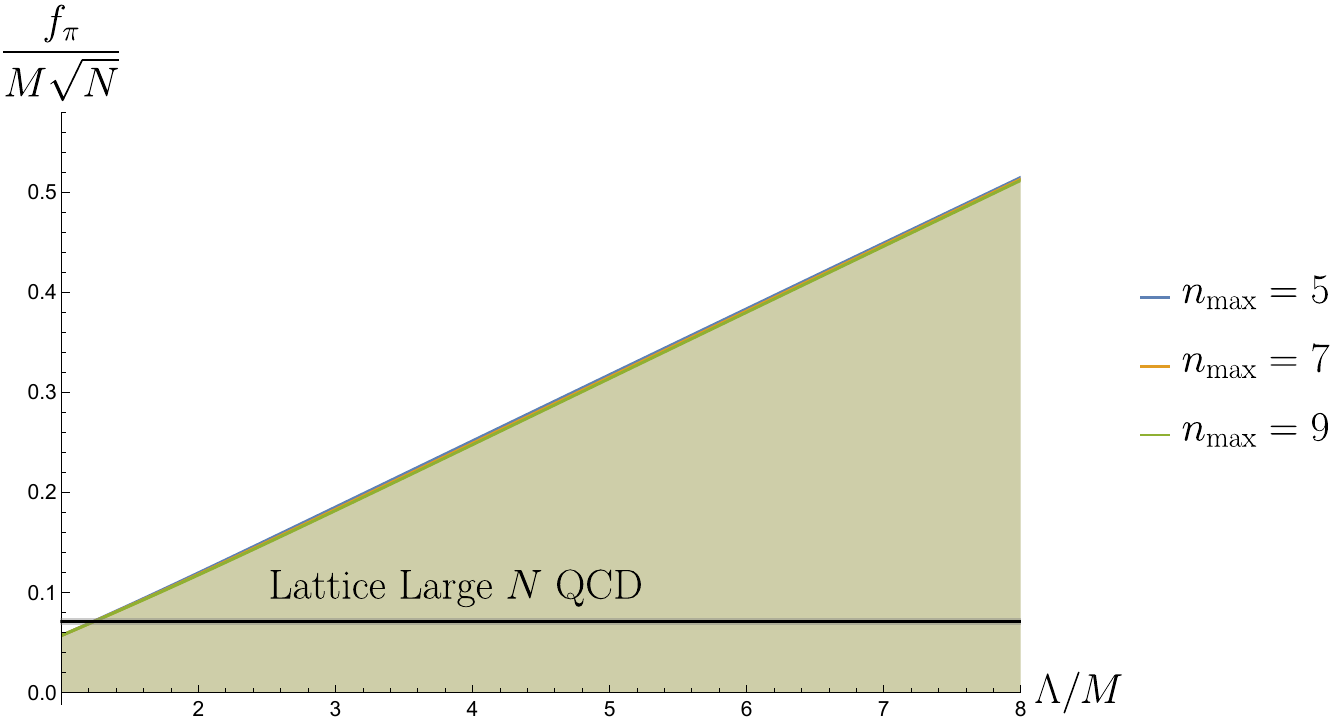}
    \caption{Bound on the ratio $f_\pi/(M \sqrt{N})$ as a function of $\Lambda/{M}$ for various values of $n_{\text{max}}$. The horizontal line marks the large $N$ lattice result \eqref{eq:fpiValue} from \cite{Bonanno:2025hzr}.}
    \label{fig:boundonfpi}
\end{figure}

Plugging in the physical definitions of each coupling in the ratio, we get
\begin{equation}\label{eq:b0fromfpi}
    \tilde b_0^{\text{SVZ}} = 4\sqrt{3}\pi \frac{f_\pi}{M\sqrt{N}}\frac{M}{\Lambda}\,.
\end{equation}
With a change of variables, we can turn the bound in figure \ref{fig:boundonb0} into an upper bound on $f_\pi/M\sqrt{N}$, shown in figure \ref{fig:boundonfpi}. Recall that $M$, the EFT cutoff, is nothing but the mass of the rho meson. So this is a bound on $f_\pi /\sqrt{N}$ measured in units of $m_\rho$!

\subsubsection{Bound on the perturbative scale}\label{sec:Pert-scale-oneloop}
This ratio remains finite in the $N\to \infty$ limit, and it has been measured in lattice simulations. The latest results use the twisted Eguchi-Kawai (TEK) model \cite{Eguchi:1982nm,Gonzalez-Arroyo:1982hwr,Gonzalez-Arroyo:1982hyq,Gonzalez-Arroyo:2010omx}, based on very few lattice sites, which lets one reach very high values of $N$. In \cite{Bonanno:2025hzr}, using up to $N=841$, they determined the following result in the chiral and continuum limit,\footnote{Here we used the large $N$ results for $f_\pi$ and $m_\rho$ reported in \cite{Bonanno:2025hzr} in units of the string tension.}
\begin{equation}\label{eq:fpiValue}
    \frac{f_\pi}{m_\rho\sqrt{N}}\simeq 0.0713 \pm 0.0034\,. 
\end{equation}
This value is marked with a horizontal black line in figure \ref{fig:boundonfpi}, with a gray band for its uncertainty. We see that it is in the allowed region for large enough values of $\Lambda$, but it is eventually excluded. We can turn this around and read the point where the lines intersect as a lower bound on the perturbative scale,
\begin{equation}\label{eq:LambdaBound}
    \frac{\Lambda}{m_\rho} \geq 1.24 \pm 0.06 \; .
\end{equation}
That is, assuming that leading order perturbation theory is valid for too low a $\Lambda$ leads to an inconsistency, and ends up ruling out QCD.

Taking the experimental real-world mass of the rho, $m_\rho \simeq 775\,$MeV, for illustration, we find ${\Lambda \gtrsim 960\, \text{MeV}}$; a remarkably low scale. Of course, this does not imply that the one-loop result can actually be trusted down to this scale, but it is surprising that one does not run into trouble earlier. In the original SVZ papers \cite{Shifman:1978bx,Shifman:1978by}, and virtually all of their followups, this scale is taken around the range $\Lambda \sim 1\,$GeV$\,-\,1.5\,$GeV. This is between the rho and the next $1^{--}$ meson, so a typical treatment makes an ansatz for the spectral density $\rho^{(\gamma\gamma)}(m^2)$ composed of a delta function for the rho (and its flavor partners $\omega,\phi$) and a continuum after $\Lambda$ given by the perturbative result \eqref{eq:asymprho} (see appendix \ref{appendix:svz} for a review). One might have expected the bootstrap to rule out such a blatant choice but, as it turns out, it does~not.

In fact, SVZ analyses with such an ansatz lead to surprisingly good estimates for the mass and coupling of the low-lying mesons. The fact that these methods give better results than they had any right to can be traced back to the fact that the spectral density $\rho^{(\gamma\gamma)}(m^2)$ of real-world QCD is actually not far from said ansatz. Indeed, $\rho^{(\gamma\gamma)}(m^2)$ is proportional to the total (inclusive) cross section $\sigma(e^+ e^-\to \text{hadrons})$,\footnote{The reason behind this is that the leading channel in $e^+e^-$ is an off-shell photon, which couples to the electromagnetic part of $J_V$. The optical theorem then relates the total cross section to the imaginary part of the two-point function of the currents.} which has been measured in electron-positron collisions since the 1970s. See e.g.\ figure 8 in \cite{Davier:2019can} for a recent report. One can see sharp peaks for the lowest-lying mesons and the early onset of the perturbative behavior! So, in retrospect, the assumption that $\Pi_{\text{pQCD}}(s)$ is a good approximation of the full $\Pi(s)$ down to very few GeVs might not be that outrageous after all (perhaps even at large $N$).

\subsubsection{Higher-loop effects}\label{sec:HigherLoop}
Some words of caution about the above results are due. Recall that the SVZ sum rules \eqref{eq:bkSVZ}, \eqref{eq:a-1SVZ} carry an associated error which depends on $\Lambda$. The leading contribution to the error of the $\tilde b_0^\text{SVZ}$ bound comes from perturbative corrections to $\Pi_{\text{pQCD}}(s)$, which enter at order $O(1/\log\Lambda^2)$.
This translates into an $O(\Lambda/\log\Lambda^2)$ error for the bound on $f_\pi/M\sqrt{N}$ of figure~\ref{fig:boundonfpi}.
To gain some insight into the actual impact of this correction on the bounds, we can repeat the above analysis with $\Pi_{\text{pQCD}}(s)$ computed at one higher loop order.

The two-loop result, originally computed for QED in \cite{Kallen:1955fb}, is now a standard result in the QCD literature, see e.g.\ \cite{Chetyrkin:1996cf}. In 't~Hooft's large $N$ limit (i.e.\ $N\to \infty$ with $\lambda \equiv 4\pi\alpha_s N$ order one), and $\overline{\text{MS}}$ scheme, it is given by 
\begin{equation}\label{eq:2loop}
    \Pi_{\text{pQCD}}(s) \simeq -\frac{N}{24\pi^2}\left(1+\frac{3 \lambda(\mu)}{32\pi^2}\right)\log(-s/\mu^2)+\cdots \,.
\end{equation}
The 't~Hooft coupling $\lambda(\mu)$ runs with energy. At high energies, we can work out its running with the one-loop beta function \cite{Gross:1973id,Politzer:1973fx}, 
\begin{equation}
    \beta(\lambda) \equiv \mu^2 \frac{d\lambda}{d\mu^2} = -\frac{11}{48\pi^2} \lambda^2\,.
\end{equation}
Solving the differential equation gives
\begin{equation}
    \lambda(\mu) = \frac{1}{\lambda(\mu_0)^{-1} + \frac{11}{48\pi^2} \log(\mu^2/\mu_0^2)}\equiv \frac{48\pi^2}{11\log(\mu^2/\mu^2_B)}\, ,
\end{equation}
where the boundary condition $\lambda(\mu_0)$, or equivalently $\mu^2_B$, should be determined from large $N$ lattice simulations in the $\overline{\text{MS}}$ scheme. In the t'Hooft limit, quark loops are suppressed by powers of $1/N$, so the running of $\lambda$ in QCD coincides with the pure Yang-Mills running. The scale $\mu^2_B$ for large $N$ Yang-Mills has been measured in the literature  \cite{Allton:2008ty,Lohmayer:2012ue,butti2023testingasymptoticscalingyangmills,Gonzalez_Arroyo_2013}.\footnote{The scheme-dependent scale $\mu_B$ is conventionally denoted as $\Lambda_{\overline{\text{MS}}}$ in the literature, here we avoid using $\Lambda$ to prevent any confusions with the perturbative scale.} In \cite{butti2023testingasymptoticscalingyangmills}, the authors use the TEK model, mentioned briefly in the previous section, to measure $\mu_B$ in units of the string tension. Combining their results with \cite{Bonanno:2025hzr} to convert to units of the rho mass yields
\begin{equation}
    \frac{\mu_B}{m_\rho}=0.30 \pm 0.01 \, , \quad \mu_B\sim 230 \text{ MeV} \,.
\end{equation}

We can now plug \eqref{eq:2loop} into the high-energy arc $|s| \sim \Lambda^2$ of SVZ dispersion relations. In doing so, one should use the 't~Hooft coupling evaluated at the characteristic scale of the process, i.e.\ $\lambda(\Lambda)$.\footnote{This is not an exact requirement. Any value around that scale would suffice, as the variation in the coupling would be of order $\lambda(\Lambda)^2$. In practical phenomenological applications, the scale is varied between $2\Lambda$ and $\Lambda/2$ to provide a crude estimate of the error from neglecting higher-loop corrections. For simplicity, we will report only the central value.} This suppresses large ``leading logs'' in the perturbative expansion by effectively resumming them.

\begin{figure}
    \centering
    \includegraphics[width=0.85\linewidth]{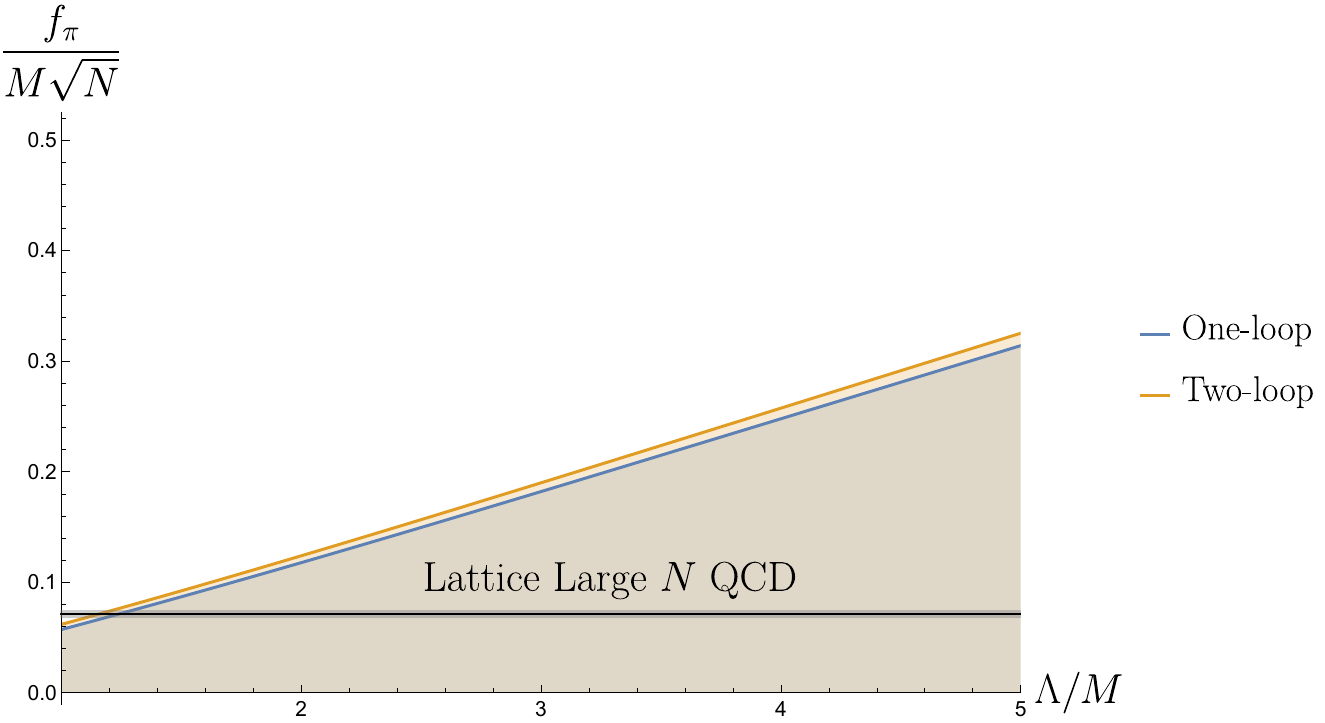}
    \caption{Bound on $f_\pi/(M\sqrt N)$ obtained from $\Pi_{\text{pQCD}}(s)$ evaluated at one loop (blue) and two loops (orange) with ${n_{\text{max}}=9}$.}
    \label{fig:fpi-twoloop}
\end{figure}

With the two-loop correction, the two-point function coefficient becomes
\begin{equation}\label{eq:two-loop-a1}
    a_{-1} = \frac{N}{24\pi^2}\Lambda^{2}\left(1+\frac{9}{22\log\left(\Lambda^2/\mu_{B}^2\right)} \right)\,,
\end{equation}
but its sum rule is still given by the right-hand side of \eqref{eq:a-1SVZ}. The numerical bound on $\tilde b_0^\text{SVZ}$ from figure \ref{fig:boundonb0} remains unchanged, but its relation to $f_\pi/\sqrt{N}$ gets modified. Inverting \eqref{eq:b0SVZdef} with the new value of $a_{-1}$ produces the orange curve in Figure \ref{fig:fpi-twoloop}. Overall, the correction does not alter the bound very significantly: the lattice value of $f_\pi/\sqrt{N}$ remains largely within the allowed region, and the lower bound on $\Lambda$ decreases only slightly, from $\sim 960$~MeV to $\sim 900$~MeV. So, while it is true that these bounds are only approximate, they are still meaningful, especially for phenomenological applications. This analysis can be systematically continued to higher (perturbative and non-perturbative) corrections.

\subsection{Bound on the pion charge radius}
Let us resume the derivation of new bounds for the low-energy form factor coefficients \eqref{eq:lowEexp} using the new SVZ-like sum rules. Next in line after $b_0$ is $b_1$. This coefficient, which is not fixed by symmetry requirements, is physically very interesting. It is traditionally parametrized in terms of a quantity dubbed the \textit{pion charge radius},
\begin{equation}
    b_1 \equiv \frac{1}{6} \langle r_{\pi}^2\rangle\,,
\end{equation}
due to its interpretation, in a certain frame, as a measure of the size of the electric charge distribution of a pion.\footnote{In the Breit frame, $(p_1+p_2)^0=0$, the form factor is represented as the Fourier transform $F(-\vec q^2)\sim \int d^3\vec x\, e^{i\vec q \cdot \vec x}\rho(\vec x)\,,$ of what resembles a non-relativistic density for the spatial charge distribution of a pion $\langle \pi|J^0_V(\vec x)|\pi\rangle$. Its expansion coefficients then correspond to moments of said density, hence the name ``form'' factor, as it encodes the shape of the pion.
}
In section \ref{sec:exclusionPlot}, we already derived a bound for this coupling normalized by $\sqrt{a_1 g_{1,0}}$, recall \eqref{eq:boundonb1}. Unfortunately, as far as we know, $a_1$ has neither been measured in experiment nor in lattice simulations, so we cannot compare that bound to data. The situation is much better with SVZ sum rules, as we can now use known quantities, such as $a_{-1}$, to normalize $b_1$.

Running the semidefinite program for a bound on
\begin{equation}
    \tilde b_1^\text{SVZ} \equiv 
    \frac{b_1}{\sqrt{a_{-1} g_{1,0}}} =
    \frac{2\pi}{\sqrt{3}}
    \frac{\langle r_{\pi}^2\rangle}{\Lambda}\frac{f_\pi}{\sqrt{N}}\,,
\end{equation}
as a function of $\Lambda/M$, without further input, produces a constant bound,
\begin{equation}
    \tilde{b}_1^{\text{SVZ}} \leq 1.246... \, , \quad \forall \Lambda/M \in [1,\infty)\,.
\end{equation}
This bound is saturated by the $su$ \textit{-pole} theory discussed in \ref{sec:exclusionPlot}: an infinite tower of mesons with increasing spin, localized at $m^2=M^2$. This theory remains an extremal solution no matter how small the $[M^2,\Lambda^2]$ window is; therefore, dialing $\Lambda$ not improve the bound.

\begin{figure}
    \centering
    \includegraphics[width=0.8\linewidth]{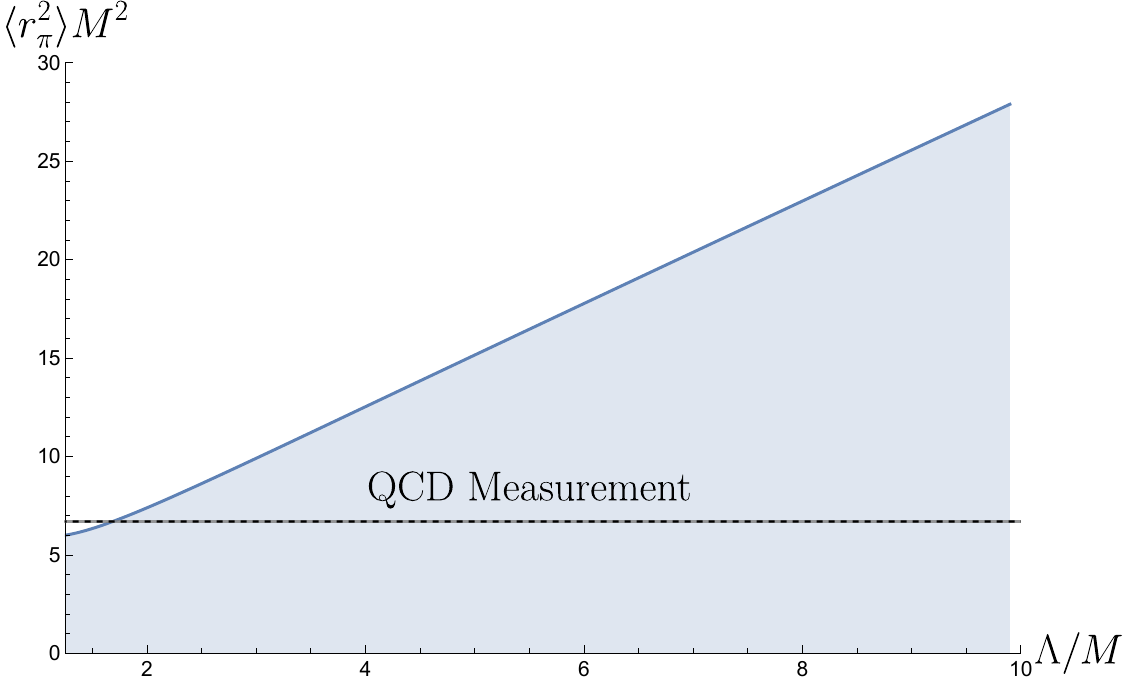}
    \caption{Bound on the pion charge radius through $\tilde{b}_1^{SVZ}$ plotted as a function of $\Lambda/M$ with $n_{\text{max}}=9$. The dashed line marks the experimental measurement of $\langle r_\pi^2\rangle m^2_\rho$.}
    \label{fig:pion-radius-bound}
\end{figure}

To introduce dependence on $\Lambda/M$, we can input the large $N$ lattice result for $f_\pi/\sqrt{N}$ through the coupling $\tilde{b}_0^{\text{SVZ}}$. This produces an increasing bound on the charge radius shown in figure \ref{fig:pion-radius-bound}.
The same words of caution as in the previous section apply here; the bound carries an error associated to the higher loop truncation.
From \eqref{eq:Flow}, we see that this is also a bound on the Wilson coefficient $\kappa_3$,\footnote{This coupling is more commonly denoted $L_9$ in the literature, when mass terms and finite $N$ corrections are included in the chiral Lagrangian.} since at large $N$, $\langle r_{\pi}^2\rangle = 12 \kappa_3/f_\pi^2$. This is a new coefficient which we could not access before in pion scattering \cite{Albert:2022oes} nor with a pion-photon mixed system \cite{Albert:2023jtd}. We are not aware of large $N$ determinations of this coupling, but we can compare the bound in figure \ref{fig:pion-radius-bound} to experimental results for real-world QCD. Rather than the Wilson coefficient, which is determined at finite $N$ by higher-loop chiral perturbation theory, we can compare directly with the experimental value for the charge radius. 

The latest PDG update \cite{ParticleDataGroup:2024cfk} reports
\begin{equation}
    \langle r_\pi^2 \rangle m_\rho^2 \simeq 6.703 \pm 0.082\,.
\end{equation}
This is marked with a dashed line (of negligible uncertainty) in figure \ref{fig:pion-radius-bound}. We see that the experimental result is allowed for sufficiently large $\Lambda/M$, but is excluded once $\Lambda$ drops below $\sim 1.7 \; m_\rho$. This significantly improves the lower bound on the perturbative scale extracted from $f_\pi/\sqrt{N}$ alone. However, it should be taken with a grain of salt because the experimental value is measured at finite $N$, and the bound weakens as higher-loop corrections are included. It would be  straightforward to include higher loop effects, but a more honest comparison would require a large $N$ lattice determination of the EFT coupling $b_1$, which is lacking as far as we are aware.

\subsection{Asymptotic freedom in pion scattering}
The --now standard-- positivity bounds on large $N$ pion scattering \cite{Albert:2022oes,Albert:2023jtd,Albert:2023seb,Fernandez:2022kzi,Ma:2023vgc,Li:2023qzs,Eckner:2024pqt,Berman:2024kdh} apply to very generic theories of weakly-coupled mesons. They do not incorporate any quantitative input from the underlying UV QCD Lagrangian (beyond large $N$ kinematics).
The SVZ sum rules that we have been discussing do; they incorporate the assumption of asymptotic freedom. It is natural to ask whether this input in the form factor system ``backreacts'' on the scattering amplitude, thereby shrinking the original plots. In general, this is not the case because we can always decouple the form factor and two-point function from the amplitude by setting $f_n=0$ in \eqref{eq:1--rhos} for any spin-one exchange in the amplitude. But we can force the systems to couple by fixing the value of a (normalized) low-energy form factor coefficient.

\begin{figure}
    \centering
    \includegraphics[width=0.9\linewidth]{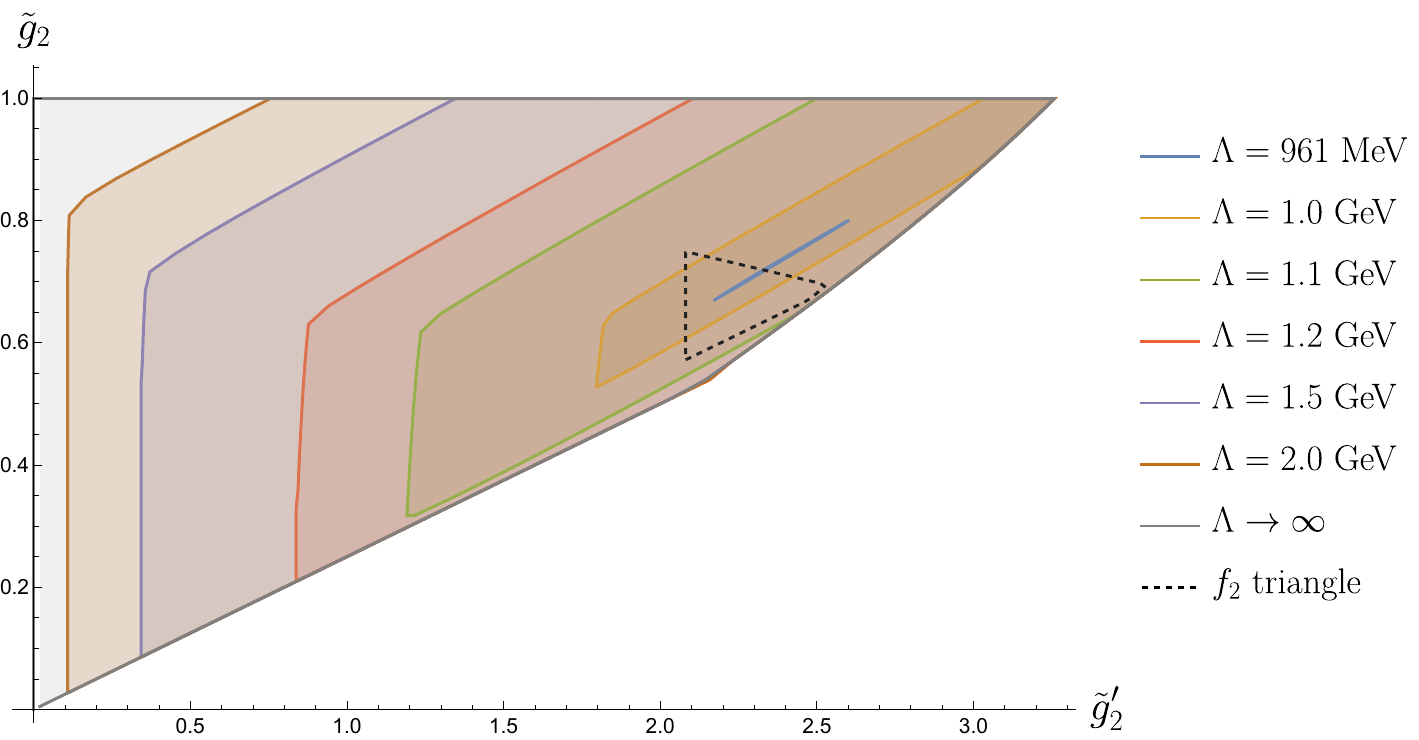}
    \caption{Regions of EFT couplings that allow for lattice QCD values at different $\Lambda$ scales, slightly higher than $\Lambda=961$ MeV, with $n_\text{max}=9$.} 
    \label{fig:shrunk-eft}
\end{figure}

The natural choice is $\tilde b_0^\text{SVZ}$, directly related to $f_\pi/\sqrt{N}$ by \eqref{eq:b0fromfpi}, which we have seen is measured with some precision in large $N$ lattice simulations. We can get the backreaction on the main exclusion plot from \cite{Albert:2022oes} as follows. We set up a semidefinite programming problem to place bounds on the three-dimensional space parametrized by $g_{2,0},g_{2,1}$ (normalized by $g_{1,0}$) and $\tilde b_0^\text{SVZ}$, and we restrict to the slice of $\tilde b_0^\text{SVZ}$ selected by the lattice result \eqref{eq:fpiValue}. We can do this at any $\Lambda$ above the lower bound \eqref{eq:LambdaBound}. The results for various values of $\Lambda$ are reported in figure \ref{fig:shrunk-eft}. We see that the assumption of asymptotic freedom in the form factor system, together with one lattice result, does constrain further the space of EFTs, thereby shrinking the allowed region.

Some comments are in order. First, the same caveats about higher-order effects discussed above apply to these bounds. They have a $\Lambda$-dependent error introduced by truncating the high-energy expansions of the two-point function and form factor. This can systematically be improved upon by including higher terms. A natural future direction would be to repeat our analysis with the two-loop result for $\Pi_{\text{pQCD}}(s)$, and or to include the first non-perturbative power correction from the condensates discussed in \eqref{eq:pOPE}. In spite of this error, one should not disregard these bounds. Indeed, in light of the small effect from higher loops showcased in figure~\ref{fig:fpi-twoloop}, we expect them to be quite accurate. So, while one should not trust their precise numerical values, they certainly point towards the correct ballpark.

Second, looking at the bounds, we see that as we decrease $\Lambda$, the allowed region migrates towards the upper-right corner of the exclusion plot. In particular, theories with a heavy contribution from scalar exchanges (living close the vertical axis) get quickly excluded. They are not compatible with an asymptotically free behavior (kicking in at low-enough energies). A similar fact has recently been observed in \cite{Cheung:2026lpv} where, using a simple system of higher-point amplitudes, they derived a bound for $\tilde g_2'$ in terms of the coefficient of the WZW term, which is linked to an anomaly and thus fixed by the UV theory.

Third, as $\Lambda$ gets down to $\sim 1\,$GeV, the bound zooms in on the region where we expect QCD to live in! This expectation comes from an exploration done in \cite{Albert:2023seb}, where bootstrap bounds where ran with the masses and couplings of the first two mesons on the leading Regge trajectory (rho and $f_2$) fixed to their real-world values, but with no UV assumptions. This produces the dashed triangle in figure \ref{fig:shrunk-eft}, which overlaps significantly with the low-$\Lambda$ form factor bounds. While neither of these bounds provide sharp rigorous predictions for large $N$ QCD, we find it very remarkable that they agree. This is one more instance where we find that asymptotic freedom in the UV reproduces the effects of the low-lying mesons in the IR, as repeatedly observed in SVZ games.

Last, for $\Lambda$ just above its lower bound compatible with the lattice measurement of $f_\pi/\sqrt{N}$, the region shrinks to a thin sliver that still overlaps (partially) with the $f_2$ triangle. One might be tempted to speculate that large $N$ QCD sits on this line, but that is probably too far fetched, as it would mean trusting the one-loop result down to very low energies. But this makes it all the more surprising that UV and IR expectations remain compatible all the way. From the bootstrap perspective, there was no reason for the bounds with two fixed low-lying mesons to agree with those with fixed asymptotic spectral densities and the pion decay constant.

\section{Conclusions and outlook} \label{sec:conclusions}
In this paper we have extended the program of carving out the space of consistent large $N$ gauge theories, initiated in \cite{Albert:2022oes}, to a bootstrap system involving the pion form factor $F(s)$ and the vector current two-point function $\Pi(s)$. The motivation for considering correlators of these local probes is that, unlike scattering amplitudes in the Regge limit, their high-energy behavior is directly controlled by the microscopic QCD Lagrangian.

Inputing quantitative information of the underlying gauge theory into the bootstrap has remained an open challenge for this program. An attempt to address this was made in~\cite{Albert:2023jtd} (see also \cite{Ma:2023vgc,Cheung:2026lpv}), leveraging on the chiral anomaly, which fixes the coefficient of the WZW term in the IR by matching to the UV theory. The approach there was to consider a mixed system of $2\to 2$ scattering amplitudes of pions and probe photons. While it led to many new technical developments, all the bounds using UV information depended on new unknown low-energy couplings. In the present paper, UV information was not introduced in the IR through anomaly matching. Rather, it entered directly in dispersion relations when probing the UV limit of the two-point function and form factor. This strategy was exploited in \cite{He:2023lyy,He:2024nwd,He:2025gws} in the context of the (finite $N$) non-perturbative $S$-matrix bootstrap.

We obtained bounds of two kinds. First, by killing the arcs at infinity, we derived rigorous bounds for low-energy form factor and two-point function coefficients. While the bounds are new, they carve out a very familiar space of amplitudes, saturated by the pervasive unphysical solutions to crossing. This was to be expected, as $F(s)$ and $\Pi(s)$ do not introduce any additional null constraints (from crossing or otherwise).

Second, by keeping the UV arcs at a large but finite scale $\Lambda$, and truncating the high-energy expansions of $F(s)$ and $\Pi(s)$, we were able to do much better. The price to pay was introducing an error to the bounds controlled by $1/\Lambda$, making them interesting from a more phenomenological perspective. This enabled us to bound new observables, such as the pion charge radius and, importantly, the pion decay constant $f_\pi$ (normalized by $\sqrt{N}$ and the mass of the rho). The bound on this ``dimension six'' operator comes as a pleasant surprise of the form factor system, as standard positivity bounds are blind to operators below ``dimension eight''. The cherry on top comes from overlaying this bound on the previous allowed space for pion amplitudes, and feeding in its large $N$ lattice value. As the perturbative scale $\Lambda$ is decreased, the allowed space zooms in towards the region where we expect QCD to sit! While these bounds are not set in stone, we take them as nontrivial evidence that the asymptotic freedom of QCD is intricately woven into the low-energy data of the pion Lagrangian.

In spite of the simplicity of the present bootstrap system, there are several avenues for its future exploration. To begin with, a very natural direction would be to extend our analysis to include higher-order corrections to the UV behaviors. These come in different guises. For the two-point function, there are perturbative and non-perturbative corrections. The former are straightforward to include, as we illustrated in section \ref{sec:HigherLoop}. The latter depend on the one-point functions $\langle O_k\rangle$ entering the OPE \eqref{eq:OOOPE}. One could take different approaches for these observables: either input them from lattice measurements or, alternatively, target them with the bootstrap and bound them. Likewise, the UV corrections for the pion form factor depend on the non-perturbative moments of the pion distribution amplitude $\phi_\pi$ \eqref{eq:light-ray}. One could take a similar course of action for these.

Another natural direction would be to repeat this story with other local probes of the system. In principle, there is no obstruction to taking any other operator $\mathcal O_i(x)$ of the UV theory and considering its two-point function and pion form factor. The vector current $J_V^\mu$ is particularly well behaved because its dimension is protected under RG flow, so it can be sharply mapped to the IR Lagrangian. A generic operator would run and flow to a complicated combination of operators in the IR, obscuring the meaning of the corresponding low-energy couplings. But, as far as the spectrum and on-shell data are concerned, the bootstrap problem seems well posed.

A candidate that is morally closer to $J_V^\mu$ is the stress tensor $T_{\mu\nu}$: it is also stable under RG flows and so it has a clear meaning in the IR. The catch is that, apart from a quark bilinear, $T_{\mu\nu}$ contains a pure glue piece.
This part couples to glueballs with a coupling enhanced by a factor of $\sqrt{N}$ compared to the meson coupling. As a result, a stress-tensor bootstrap will necessarily require extending the system to the glueball sector. This would still be very interesting.

Finally, and perhaps needless to mention, there is ``the more the merrier'' direction. Systems closed under positivity and crossing are not mutually exclusive. On the contrary, one can continue to enlarge mixed bootstrap systems. For instance, once rho meson scattering is under control \cite{rhos}, one can envision a large bootstrap problem involving pions, rhos and their corresponding vector form factors. This direction keeps proving useful in the conformal bootstrap, see e.g.\ \cite{Chang:2024whx}, where a system involving stress tensors has led to a determination of the the Ising model scaling dimensions of unprecedented precision. We optimistically anticipate similar progress in the current problem. With the continued collective effort of the community, the dream of cornering large $N$ QCD keeps getting closer.

\acknowledgments
We thank Simon Caron-Huot, Miguel Correia, Gabriel Cuomo, Matthew Forslund, Johan Henriksson, Alexandre Homrich, Martin Kruczenski, Yue-Zhou Li, Ian Moult, Julio Parra-Martinez, Bruno Scheihing-Hitschfeld, George Sterman and Alessandro Vichi for useful discussions and suggestions.
The authors are grateful to the Aspen Center for Physics (supported by NSF
grant no. PHY-2210452) and ICTP-SAIFR (supported
by FAPESP grant 2021/14335-0), where parts of this
work were completed, for their hospitality.
JA acknowledges partial support from Simons Foundation grant 917464 (Simons Collaboration on Confinement and QCD Strings).
The work of LR and DK is supported in part  by the NSF grant PHY-2513893 and by the Simons Foundation
grant 681267 (Simons Investigator Award).

\appendix

\section{Traditional SVZ approach}\label{appendix:svz}
In this appendix, we provide a brief pedagogical review of the traditional sum rules introduced by Shifman, Vainshtein, and Zakharov \cite{Shifman:1978bx,Shifman:1978by} for the benefit of the reader unfamiliar with the subject. For a more detailed treatment and a survey of modern applications, we refer the reader to the reviews \cite{Reinders:1984sr,Colangelo:2000dp,Gubler_2019}.

As we discuss in the main text, the SVZ approach relates two descriptions of the same correlator in different regimes. At short distances, the correlator admits an OPE expansion whose Wilson coefficients can be computed in perturbation theory, and the same correlator is fixed by its physical spectrum through a dispersion relation. Equating these two representations, after suitable manipulations, yields constraints on hadronic data in terms of short-distance QCD input.

$\Pi(s)$ is analytic in the complex $s$ plane away from the positive real axis, with discontinuity
$\mathrm{Im}\,\Pi(s)=\pi \rho^{(\gamma\gamma)}(s)$ for $s>0$. For vector currents, one subtraction is enough to write down the dispersion relations; in a more general setup where UV behavior requires $k$-subtractions, the dispersion relations take the following form
\begin{equation}\label{eq:svzdispersion}
\Pi(s)= \frac{s^k}{\pi}\int_{s_0}^{\infty} ds'
\frac{\text{Im}\Pi(s')}{s'^k\left(s'-s\right)} + \sum_{l=0}^{k-1}a_l s^l \; ,
\end{equation}
where the coefficients $a_l$ are the derivatives of the propagator evaluated at $s'=0$. For vector currents, $k=1$, and $a_0$ corresponds to the EFT coupling $\Pi(0)$. Up to this point, no SVZ-specific assumption has been made, and \eqref{eq:svzdispersion} follows from analyticity and subtractions.

To get rid of the subtraction terms, SVZ perform a Borel transform, which moreover improves the convergence of the perturbative expansion as we now explain. Formally, the Borel transform is defined as
\begin{equation}\label{eq:borel_def}
\mathcal{B}_{M_B^2}\!\left[f(q^2)\right]
\equiv
\lim_{\substack{q^2,n\to\infty\\ q^2/n=M_B^2}}
\frac{(q^2)^n}{(n-1)!}\left(-\frac{d}{dq^2}\right)^n f(q^2)\, ,
\end{equation}
with parameter $M_B^2>0$. The Borel transform of any polynomial in $q^2$ vanishes, and one can check that 
\begin{align}
    \mathcal{B}_{M_B^2}\!\left[\frac{1}{s+q^2}\right] &= \lim_{\substack{q^2,n\to\infty\\ q^2/n=M_B^2}} \frac{1}{M_B^2}\left(1+\frac{s}{nM_B^2}\right)^{-(n+1)}  =\frac{1}{M_B^2}\exp\left(-\frac{s}{M_B^2}\right) \,, \\
     \mathcal{B}_{M_B^2}\!\left[\frac{(-q^2)^k}{s+q^2}\right]&=(-1)^k\left( \mathcal{B}_{M_B^2}\!\left[\sum_{l=0}^{k-1}(-s)^l(q^2)^{k-l-1}\right]+ \mathcal{B}_{M_B^2}\!\left[\frac{(-s)^k}{s+q^2}\right]\right)=\frac{s^k}{M_B^2}\exp\left(-\frac{s}{M_B^2}\right) \, .\nonumber
\end{align}
Applying $\mathcal{B}_{M_B^2}$ to the dispersion relation in $\eqref{eq:svzdispersion}$, the subtraction terms drop, leaving the Borelized sum rule,
\begin{equation}\label{eq:borelizedsumrule}
      \mathcal{B}_{M_B^2}\!\left[\Pi(q^2)\right] = \frac{1}{\pi M_B^2}\int_{s_0}^\infty ds' \; \text{Im}\, \Pi(s') \exp\left(-\frac{s'}{M_B^2}\right)\,.
\end{equation}

The idea is to evaluate the left hand side of $\eqref{eq:borelizedsumrule}$ using the OPE expansion in the far Euclidean region $-s=q^2>0$ where no singularities are present, and equate it to the hadronic spectral function on the right.
The first few terms contributing to the OPE were given in eq. $\eqref{eq:pOPE}$. The Borel transform of the power series improves the convergence of the OPE by the simple result,
\begin{equation}
  \mathcal{B}_{M_B^2}\!\left[\frac{1}{(q^2)^k}\right] = \frac{1}{(k-1)!}\frac{1}{M_B^{2k}}\,,
\qquad (k=1,2,\ldots)\,
\end{equation}
and the terms at higher order are factorially suppressed. Comparing the Borelized sum rule $\eqref{eq:borelizedsumrule}$ to the original dispersion relation $\eqref{eq:svzdispersion}$, the subtraction terms are automatically killed off, the convergence of the OPE is improved, and the higher excited states in the spectral density are exponentially suppressed.

At this point, SVZ inputs spectral assumptions to derive relations among couplings. In the large-$N$ setup, the spectral density is meromorphic, and we can safely separate the first excitation,
\begin{equation}
\rho^{(\gamma\gamma)}(s) = f_\rho^2\,\delta\!\left(s-m_\rho^2\right) + \rho^{(\gamma\gamma)}_{\text{cont}}(s)\theta(s-s_{\text{cont}})\, .
\end{equation}
For finite $N$, this is the narrow resonance approximation. To match the literature, we now restrict to the real-world value $N=3$. Inserting the ingredients into the sum rule yields
\begin{equation}\label{eq:svz-integral}
    \frac{1}{8\pi^2}\left(1+\frac{\alpha_s}{\pi}\right)+\frac{\alpha_s}{24 \pi M_B^{4}} \langle G^a_{\rho\sigma} G^{a \, \rho\sigma}\rangle = \frac{f_\rho^2}{M_B^2}e^{-\frac{m_\rho^2}{M_B^2}}  +  \int_{s_{\text{cont}}}^\infty ds' \frac{\rho^{(\gamma\gamma)}_{\text{cont}}(s')}{M_B^2}\exp\left(-\frac{s'}{M_B^2}\right) \,,
\end{equation}
where the identity contribution is the two-loop result at finite $N=3$, and we have only kept up to $\mathcal{O}(s^{-2})$ terms in the OPE.

The final assumption is to input quark-hadron duality to the continuum above the threshold; that is, that $\rho^{(\gamma\gamma)}_{\text{cont}}$ is well approximated by the two-loop perturbative result. The integral can easily be carried out, which leads to the celebrated sum rule for $f_\rho^2$,
\begin{equation}\label{eq:svzsumrule}
    \frac{M_B^2 }{8\pi^2}\left(1+\frac{\alpha_s}{\pi}\right)\left(1- e^{-\frac{s_{\text{cont}}}{M_B^2}}\right) + \frac{\alpha_s }{24 \pi M_B^{2}} \langle G^a_{\rho\sigma} G^{a \, \rho\sigma}\rangle = f_\rho^2 e^{-\frac{m_\rho^2}{M_B^2}} \;.
\end{equation}
The above sum rule relates the hadronic parameters $f_\rho$ and $m_\rho$ to the QCD input $\alpha_s$ and $\langle \text{Tr} \, GG \rangle$, alongside the auxiliary Borel parameter $M_B^2$. In practice, one takes the logarithm of $\eqref{eq:svzsumrule}$ and differentiates with respect to $M_B^{-2}$, to isolate $m_\rho^2$ on the right-hand side and obtain a second relation. However, both relations depend on the nonphysical $M_B^2$; hence, one needs to settle on an appropriate range for $M_B^2$ before using the sum rule to extract $f_\rho^2$ and $m_\rho^2$.

The exponential weight inside the integral in eq. $\eqref{eq:svz-integral}$ flattens to 1 when $M_B^2$ is taken too large over a wide range of $s$, so the right-hand side is dominated by the continuum rather than the isolated resonances. On the other hand, if $M_B^2$ is too small, the power corrections to the OPE are no longer small, and the truncated expression ceases to be reliable. For these reasons, $M_B^2$ needs to be in an intermediate range where both assumptions hold, and the extracted parameters are stable; this region of $M_B^2$ values is called a \textit{Borel window}. Finding a Borel window for a given sum rule is not guaranteed a priori; for some channels, the lower bound on $M_B^2$ exceeds the upper bound. However, for the $\rho$ channel, a Borel window does indeed exist, and the relation $\eqref{eq:svzsumrule}$ predicts the measured coupling and the mass to high accuracy. We refer the reader to the reviews \cite{Reinders:1984sr,Colangelo:2000dp,Gubler_2019} for a detailed account of the determination of the Borel window and the predictive power of the sum rules.

\bibliographystyle{ytphys}
\bibliography{refs}

\end{document}